\documentclass{aa}  
\usepackage{natbib}

\usepackage{graphicx}
\usepackage{txfonts}
\usepackage{lipsum}
\usepackage{subcaption}         % necessary for continued figures, example in section 3
\usepackage{lscape}             % to rotate a single page table, example in appendix.
\usepackage{placeins}           % useful with \FloatBarrier, to keep 
\usepackage{tikz-feynman}
\usepackage{siunitx}
\usepackage{upgreek}
\usepackage{subcaption}
\usepackage{multirow}
\usepackage{relsize}
\usepackage{float}

\begin{document}

%%%%%%%%%%%%%%%%%%%%%%%%%%%%%%%%%%%%%%%%
% if you use custom commands in your title,
% ensure to check your title when submitting!
%%%%%%%%%%%%%%%%%%%%%%%%%%%%%%%%%%%%%%%%
%   \title{Calculation of Electron-Positron Pair Production from the Cosmic Photon Background as a Function of Redshift}
   \title{Cosmological $\gamma$-$\gamma$ Pair-Production Background}
%%   \subtitle{Subtitle}

%%%%%%%%%%%%%%%%%%%%%%%%%%%%%%%%%%%%%%%%
% Please separate each author with the \and command
%
% Please do not include ORCIDs next to author names.
% Only ORCIDs authenticated by individual authors in EDPS
% editorial system will be taken into account.
% ORCIDs included here will be removed.
%%%%%%%%%%%%%%%%%%%%%%%%%%%%%%%%%%%%%%%%

    \author{
        Mika\,A. Gelowicz\inst{1,2}\thanks{Corresponding author: \email{s85mgelo@uni-bonn.de}}
        \and Thomas Siegert\inst{1}\thanks{Corresponding author: \email{thomas.siegert@uni-wuerzburg.de}}
        \and Saurabh Mittal\inst{1}
        \and Laura Eisenberger\inst{1}
        \and Dimitris Tsatsis\inst{1}
        \and Niklas\,C. Bauer\inst{1}
        \and Rudi Reinhardt\inst{1}
        \and Patrik Ehrmann\inst{1}
        \and Manja\,L. Zimmerer\inst{1}
        \and Hiroki Yoneda\inst{3,4,5,6}
        \and Tomohiko Oka\inst{1}
        \and Tristan Bouchet\inst{1}
        \and Manuel\,R.\,H.\,W. Skalka\inst{1}
        }

        \authorrunning{Gelowicz et al.}

    \institute{
        Julius-Maximilians-Universität Würzburg,
        Fakultät für Physik und Astronomie,
        Institut für Theoretische Physik und Astrophysik,
        Lehrstuhl für Astronomie,
        Emil-Fischer-Str.~31,
        D-97074 Würzburg, Germany
        \and
        Rheinische Friedrich-Wilhelms-Universität Bonn, Argelander-Institut für Astronomie (AIfA), Auf dem Hügel 71, D-53121 Bonn, Germany
        \and
        The Hakubi Center for Advanced Research, Kyoto University, Yoshida Ushinomiyacho, Sakyo-ku, Kyoto 606-8501, Japan
       \and
       Department of Physics, Kyoto University, Kitashirakawa Oiwake-cho, Sakyo-ku, Kyoto 606-8502, Japan
       \and
       RIKEN Nishina Center, 2-1 Hirosawa, Wako, Saitama 351-0198, Japan
       \and
       Kavli Institute for the Physics and Mathematics of the Universe (WPI), UTIAS, The University of Tokyo, 5-1-5 Kashiwanoha, Kashiwa, Chiba 277-8583, Japan
}

   \date{Received December XX, 2025}

% \abstract{}{}{}{}{}
% 5 {} token are mandatory
 
  \abstract
  % context heading (optional)
  % {} leave it empty if necessary  
   {The origin of positrons is one of the unsolved puzzles in astrophysics as the majority of sources are still unidentified. The Cosmic Photon Background (CPB) is the isotropic radiation spanning the entire electromagnetic spectrum. Interactions of the CPB with itself may pose a promising source of positrons and secondary emission.}
  % aims heading (mandatory)
   {We aim to calculate the electron-positron pair production rate from the $\gamma$-$\gamma$ pair-production of the CPB with itself for redshifts up to $z = 10$, and determine the annihilation spectrum as well as Inverse Compton emission and bremsstrahlung.} 
  % methods heading (mandatory)
   {The CPB is decomposed into a sum of gray body functions, of which each is being evolved according to source type luminosity functions and redshift. We compute the pair-production rate by integrating the angle- and energy-dependent cross section over the evolving CPB. The pairs produced at each redshift are then propagated towards $z=0$, taking into account a cosmological, intergalactic, energy loss function. The photon emission is calculated per redshift and then line-of-sight integrated towards a contribution of the Cosmic Gamma-Ray Background (CGB) today.}
  % results heading (mandatory)
   {The resulting pair-production emissivity increases steeply from $z=0$ of about $\SI{2.0e-36}{s^{-1}\,cm^{-3}}$ to a peak of $\SI{1.8e-31}{s^{-1}\,cm^{-3}}$ at $z=2.7$, then declines again. This yields a total cosmic pair-production rate on the order of $10^{54}\,\mathrm{e^+\,s^{-1}}$ up to redshift $10$. The secondary emission of pairs experiencing Inverse Compton scattering off the CPB results in a sizable contribution to the CGB.}
  % conclusions heading (optional), leave it empty if necessary
   {The pairs from cosmological $\gamma$-$\gamma$ absorption provide a minimum level of secondary emission which needs to be taken into account for any CGB study. Especially in the range from $1${\,MeV} to {$1$\,GeV}, this background can make up 10--20\% of the total CGB emission {and may substantially reduce the gap between MeV observations and models}.}


   \maketitle
   \nolinenumbers

\section{Introduction}\label{sec:intro}
Positron-electron annihilation remains one of the most intriguing and unresolved puzzles in astrophysics, as the majority of positron sources are still unidentified.
Several promising candidates have been proposed, including nucleosynthesis via $\beta^+$-decay in stellar environments, $\gamma$-$\gamma$ pair-production in the dense radiation fields surrounding accreting compact objects, and exotic processes such as the annihilation or decay of dark matter particles like WIMPs \citep[see][for recent reviews]{Prantzos2011_511,Siegert2023}.
Pair production is typically expected to be efficient only in regions with extremely high photon densities.
However, this limitation might be misleading by the vast cosmological volumes, in which these processes may happen, albeit rarely.

The Cosmic Photon Background (CPB) describes the quasi-isotropic photon emission between $\sim 10^{-7}$\,eV and $\sim 10^{12}$\,eV \citep{Hill2018}.
The dominant emission at any redshift is the Cosmic Microwave Background (CMB), peaking at photon energies around $\sim 10^{-4}$\,eV at redshift $z=0$.
Other strong isotropic radiation fields today are the Cosmic Infrared and Cosmic Optical Background (CIB, COB).
These fields are typically used to calculate the absorption of high-energy photons beyond 100\,GeV to obtain a cutoff in the measured spectra of AGNs \citep[e.g.,][]{Dwek2013_EBL}.
However, the number of pairs produced is not dominated by the CMB/CIB and TeV/PeV emission; instead, it is dominated by the optical/UV and GeV emission.
This is the case because, while the CIB and COB photon fields are roughly compatible, the GeV emission is orders of magnitude stronger than the TeV--PeV fluxes.
But because the GeV emission is hardly suppressed, this effect is rarely taken into account.

The differential spectrum of pairs produced by $\gamma$-$\gamma$ pair-production has been calculated exactly in \citet{Boettcher1997}.
We use their work to calculate the differential emissivity of pairs as a function of redshift by propagating the different components of the CPB backwards, given the luminosity functions of the sources responsible.
%
%In this paper, we investigate the $\gamma$-$\gamma$ pair-production from interactions of the CPB with itself as a function of redshift.
%
The paper is structured as follows:
First we recapitulate on the process of $\gamma$-$\gamma$ pair-production in Sect.\,\ref{Background}.
In Sect.\,\ref{Methodology}, we describe the decomposition of the CPB into multiple black bodies with dilution factors as amplitudes (gray bodies), and how the sources responsible for the individual components evolve with redshift.
The resulting particle spectra are then propagated in Sect.\,\ref{sec:secondaries} in a cosmological diffusion loss equation with sources, sinks (annihilation), and energy losses as a function of time (redshift) as there is no steady state.
We conclude our study and provide an outlook in Sect.\,\ref{Conclusion & Outlook}.

\section{Theoretical Background}\label{Background}
\subsection{$\gamma$-$\gamma$ Pair-Production}\label{sec:pair_production}
Two interacting photons can produce an electron-positron pair if their energy is large enough, that is, above the threshold energy of $E_1E_2 \geq m_e^2$ in natural units.
The cross section for this process,
\begin{equation}
    \sigma_\mathrm{pair} = \frac{3 \sigma_T}{16} \left(1 - \beta^2\right) 
    \left[ \left(3-\beta^4\right) 
    \ln\left(\frac{1+\beta}{1-\beta} \right) - 2\beta\left(2-\beta^2\right)
 \right] \mathrm{,}
 \label{eq: cross section pair production}
\end{equation}
with $\sigma_T = 6.67 \times 10^{-25}\,\mathrm{cm^{2}}$ being the Thomson cross section, with the electron's speed $\beta$ expressed as the reduced energies of the two interacting photons, $\varepsilon_{1/2} \equiv E_{1/2}/m_e$, and their incidence angle $\theta_\mu$,
\begin{equation}
    \beta = \sqrt{1 - \frac{2}{\varepsilon_1\varepsilon_2 (1-\cos{\theta_\mu)}}}\mathrm{,}
    \label{eq: beta}
\end{equation}
peaking around $\beta = 0.7$, and reaching a maximum of $\sigma_\mathrm{pair}^{\mathrm{max}} \approx \SI{1.7e-25}{cm^{2}}$ \citep{JauchRohrlich1976, greiner2008quantum}.

Given two photon fields $n_\mathrm{ph}(\varepsilon_1)$ and $n_\mathrm{ph}(\varepsilon_2)$, we can estimate a total pair production emissivity in units of $\mathrm{cm^{-3}\,s^{-1}}$ by
\begin{equation}   
    \dot{n}_\pm(\varepsilon_1,\varepsilon_2) = c \, n_\mathrm{ph}(\varepsilon_1) \, n_\mathrm{ph}(\varepsilon_2) \, \sigma_{\mathrm{pair}}(\varepsilon_1,\varepsilon_2)\mathrm{,}
    \label{eq: naive denisty rate}
\end{equation}
where $c$ is the speed of light.
For example, for head-on collisions, we can estimate the emissivity of the COB ($\lambda \sim 550\,\mathrm{nm} \approx 1\,\mathrm{eV}$) with the CGB around 1\,TeV.
With a photon density of $n_\mathrm{COB} = 10^{-4}\, \mathrm{cm}^{-3}$ and $n_{\mathrm{CGB}} = \SI{3e14}{cm^{-3}}$, respectively \citep{longair2007galaxy}, we find $\dot{n}_\mathrm{\pm} \approx 10^{-36}\,\mathrm{cm}^{-3}\,\mathrm{s}^{-1}$ as an order of magnitude estimate, which sets the scale for this problem.
The pair production cross section drops rapidly for $\beta \lesssim 10^{-2}$ or $\beta \gtrsim 0.999$, so that there is a ``sweet-spot'' for pair production for the combination of photon energies:
While TeV--PeV photons are efficiently absorbed by CMB and COB photons (the ``extragalactic background light'', EBL), there is hardly any GeV attenuation.
However, the absorption process does not stop at GeV energies, so that the Cosmic Ultraviolet Background (CUB), together with the GeV background efficiently produces pairs because the UV background is comparable in amplitude to the optical background, and the GeV background is much stronger than the TeV--PeV background \citep{Hill2018}.

For the differential pair production spectrum, we have to take into account the different kinematic conditions, following \citet[][and references therein]{Boettcher1997}.
The injection spectrum is calculated by
\begin{equation}
    \dot{n} = \frac{c}{4} 
    \int_0^\infty \mathrm{d}\varepsilon_1\, n_\mathrm{ph}(\varepsilon_1)    \int_{\varepsilon_2^L} ^\infty 
    \mathrm{d}\varepsilon_2 \,n_\mathrm{ph}(\varepsilon_2)
    \int_{-1}^{1 - \frac{2}{\varepsilon_1\varepsilon_2}} \mathrm{d}\mu\,(1-\mu) \frac{\mathrm{d}\sigma}{\mathrm{d}\gamma} \mathrm{,}
    \label{eq:pair_spectrum}
\end{equation}
where $d\sigma/d\gamma$ is the differential pair production cross section and $\mu_\mathrm{th} = 1-2/(\varepsilon_1\varepsilon_2)$ is the threshold angle for pair production.
We refer the reader to \citet{Boettcher1997} for the details of the analytical solution of Eq.\,(\ref{eq:pair_spectrum}) until only the integrals over the photon fields are left:
\begin{equation}
    \dot{n} = \frac{3}{4} \sigma_T c \int_0^\infty \mathrm{d} \varepsilon_1\frac{n_\mathrm{ph} (\varepsilon_1)}{\varepsilon_1^2}    \int_{\varepsilon_2^L}^\infty \mathrm{d} \varepsilon_2\frac{n_\mathrm{ph}(\varepsilon_2)}{\varepsilon_2^2} \cdot \mathcal{F}(\varepsilon_1,\varepsilon_2)\mathrm{.}
    \label{eq:final_spectrum_integral}
\end{equation}
Here, $\mathcal{F}(\varepsilon_1,\varepsilon_2)$ is a lengthy but analytical function, and $\varepsilon_2^L = \mathrm{max}(\varepsilon_1^{-1},\gamma + 1 - \varepsilon_1)$ is again related to the pair production threshold.
In principle, any differential photon field can be included in Eq.\,(\ref{eq:final_spectrum_integral}), but for the purpose of this paper, we only use diluted blackbody spectra.
This has the advantage that individual components of the CPB can be propagated as a function of redshift as is required to obtain the cosmological pair injection spectrum.
We note that this process is inevitable and does have to occur as the absolute minimum injection of leptons into the intergalactic medium (IGM).

\subsection{The Cosmic Photon Background}\label{sec:CPB}
\begin{figure*}
    \centering

    \includegraphics[width=0.9\linewidth]{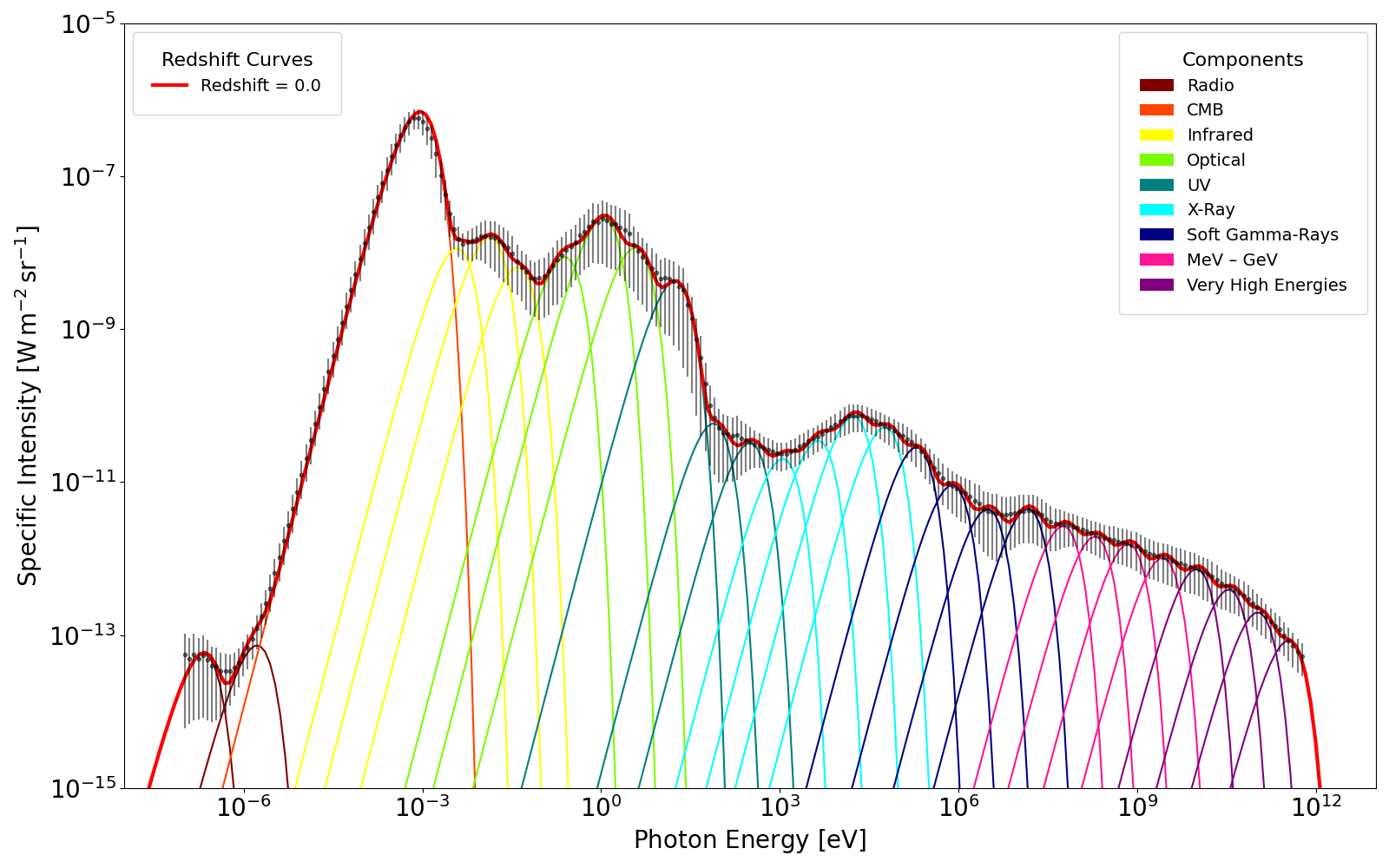}
    \caption{CPB spectrum \citep[black data points; data and uncertainties taken from][]{Hill2018} together with decomposition into blackbody functions. The colors indicate the different ranges (see Tab.\,\ref{tab: sources and ranges CPB}).
}
    \label{CPB redshift 0.0}
\end{figure*}

We treat the CPB as the isotropic radiation that ranges across the entire electromagnetic spectrum.
It is typically separated into seven different energy ranges \citep[e.g.,][]{Hill2018}, even though there are many more astrophysical components (see Sect.\,\ref{Methodology}).
We show the spectrum of the CPB from $10^7$\,Hz ($10^{-7}$\,eV) to $10^{26}$\,Hz ($10^{12}$\,eV) in Fig.\,\ref{CPB redshift 0.0}, decomposed into black bodies.
In the following, we briefly recapitulate on the typically used classification, and summarize the main sources in Tab.\,\ref{tab: sources and ranges CPB}.

\subsubsection{Cosmic Radio Background (CRB)}\label{sec:CRB}
In the radio regime, the CPB is dominated by free-free emission and synchrotron radiation, whose sources are star-forming galaxies (SFGs) and Active Galactic Nuclei (AGNs), respectively \citep{Tompkins2023}.
As free-free emission is most prevalent in H\,II regions, which are ionized by stars, diffuse radio emission is also closely linked to the star formation rate (SFR) and star formation history of the Universe \citep{Mancuso2015}.
For energies lower than $\sim 10^{-7}$\,eV, the CRB is limited by the plasma frequency of the interstellar medium (ISM), below which the Universe is virtually opaque.

\subsubsection{Cosmic Microwave Background (CMB)}\label{sec:CMB}
The CMB is perhaps the most studied part of the CPB in terms of cosmology \citep{Penzias1965_CMB}.
Its origin is typically described in the surface of last scattering at a redshift of $z \approx 1100$, when protons and electrons were able to form neutral hydrogen.
Through that process, photons that were previously scattered by the charged particles, could pass unhindered through space -- the Universe became transparent.
Because of the expansion of the Universe, this radiation, which used to have a temperature of around $\SI{3000}{K}$, gradually cooled down and can now be seen as an almost perfect blackbody radiation of $T_\mathrm{CMB,0} = \SI{2.725}{K}$ \citep[e.g.,][]{lahav2024cosmologicalparameters2023}.

\subsubsection{Cosmic Infrared Background (CIB)}\label{sec:CIB}
The CIB is an important tracer for the cosmic SFR of the Universe, as approximately $50\%$ of the total radiative energy output from stars over the history of the Universe lies in that energy range.
Additionally, dust in galaxies heated by stars and the most redshifted galaxies contribute to the CIB \citep{Lagache2005}.
Because of the strong foreground emission from the Sun, planets, and the Milky Way, it is hard to measure exactly.
Often, only upper or lower limits that rely on theoretical models and estimations are available \citep{Hill2018}.
Combining upper and lower limits on the CIB provides a reasonable band for its intensity.

\subsubsection{Cosmic Optical Background (COB)}\label{sec:COB}
The COB is in most regard similar to the CIB.
Its sources are, likewise, stars and galaxies, but in this case, the observed light originates directly from the sources, rather than being
re-emitted by dust.
Other sources are marginal and are typically neglected \citep{Hill2018}.

\subsubsection{Cosmic Ultraviolet Background (CUB)}\label{sec:CUB}
The CUB is even less understood than the two previous ranges.
Measurements are limited by the same factors and, additionally, by the UV-absorption of hydrogen and Earth's atmosphere.
Especially important are two wavelengths in that energy range: $\SI{912}{\angstrom}$ and $\SI{228}{\angstrom}$.
The former corresponds to the ionization energy of neutral hydrogen of $\SI{13.6}{eV}$, the latter to the ionization energy of singly ionized helium.
Both elements have large abundance in galaxies and models rely heavily on their properties \citep{Khaire2019}.
The CUB generally consists of galactic contributions, such as light of hot, young stars, as well as AGN-contributions in the form of quasars.

\begin{table*}
\centering
    \begin{tabular}{|l|l|l|}
    \hline
    {} & {Main sources} & {Frequency/wavelength/energy range} \\
    \hline
    CRB & Synchrotron radiation, AGNs & $\nu < 10\,\mathrm{GHz}$ \\
    CMB & Surface of last scattering & $\nu = 10 - 1000\,\mathrm{GHz}$ \\
    CIB & Dust heated by stars, old galaxies & $\lambda = 3 - 300\,\upmu\mathrm{m}$ \\
    COB & Direct starlight & $\lambda = 300 - 3000\,\mathrm{nm}$ \\
    CUB & Hot, young stars & $\lambda = 30 - 300\,\mathrm{nm}$ \\
    CXB & Accretion disks around AGNs & $E = 0.1 - 100~\mathrm{keV}$ \\
    \multirow{3}{*}{CGB} & Quasars, FSRQ, SNe \makebox[4.6cm][r]{(soft $\gamma$-ray)} & $E = 100\,\mathrm{keV} - 20\,\mathrm{MeV}$ \\
     & FSRQ, BL\,Lacs \makebox[5.3cm][r]{(MeV--GeV)} & $E = 20\,\mathrm{MeV} - 5\,\mathrm{GeV}$ \\
    & BL\,Lacs \makebox[6.5cm][r]{(VHE)} & $E > 5\,\mathrm{GeV}$ \\
    \hline
    \end{tabular}
    \caption{Main sources making up each energy range of the CPB and their corresponding photon ranges \citep{Hill2018}.}
    \label{tab: sources and ranges CPB}
\end{table*}

\subsubsection{Cosmic X-Ray Background (CXB)}\label{sec:CXB}
The observational challenges associated with the previous lower energy ranges are no longer present in the CXB, as more telescopes observing in the X-ray regime exist, photons are less absorbed by the ISM, and there is no strong foreground emission that needs to be taken into account.
For the most part, AGNs in form of quasars (QSOs) are the dominant contributors, but there can be a galactic contribution for lower energies from hot gas and accreting compact objects \citep{Shen2020}.

\subsubsection{Cosmic $\gamma$-Ray Background (CGB)}\label{sec:CGB}
The CGB makes up the largest part of the CPB in terms of photon energies, starting in the hard X-ray range and stretching into TeV energies.
Because of the wide energy range, the CGB is split up further into three subcategories:

\paragraph{Soft $\gamma$-Ray:}
Soft $\gamma$-rays are typically considered from a few 100\,keV to some MeV -- the range of nuclear physics.
Its sources are, especially in the low MeV range, not well understood and there does not seem to be a consensus on what sources precisely make up the CGB:
It appears to be dominated mainly by flat spectrum radio quasars (FSRQs) \citep{Ajello2009}, with typical quasars being strong contributors for energies of a few $\SI{100}{keV}$ \citep{Marcotulli2022}, while supernovae being somewhat co-dominant in the MeV range \citep{Ruiz-Lapuente2016}.

\paragraph{MeV -- GeV:}
This part of the CGB was previously believed to be dominated by blazars, making up about $50\%$, and by radio galaxies as well as star-forming galaxies (each making up about $20\%$) -- thus being close to explaining the complete background \citep{inoue2014cosmicgammaraybackgroundradiation}.
Later studies suggested that the contribution of galaxies cannot be more than about 10\% \citep{Fukazawa2022, Ajello_2020, chen2025diffuse}.
FSRQs and BL Lacertae objects (BL\,Lacs), appear to be contribute in roughly equal parts \citep{Ajello2012, Ajello2014}, but uncertainties still remain.
Interestingly, we show that the cosmological pair production and the subsequent Inverse Compton emission contributes tens of per cent in this range (see Sect.\,\ref{sec:secondaries}).

\paragraph{Very High Energies (VHE):}
At GeV energies, the contribution is supposedly straight-forward as their sources are purely BL\,Lacs \citep{Ajello_2015}.
For energies beyond $\SI{300}{GeV}$, observations are limited by the large cross section of these photons interacting with those comprising the CIB, leading to electron-positron pairs.

\section{Isotropic Components as a Function of Redshift}\label{Methodology}
\subsection{Photon Spectral Decomposition into Blackbodies}\label{sec:b_decomposition}
The choice of decomposing the CPB into blackbody spectra is arbitrary but is motivated by the simplicity and smoothness.
Certainly, also a power-law decomposition would work, but since the CMB is, indeed, a perfect blackbody, we use this functional form.
We used a data compilation of different observations from \citet{Hill2018} to model the CPB.
We find that using 28 blackbody components provides an adequate fit to the data, yielding small residuals without introducing unnecessary complexity.
These components were then fitted to the data set by $\chi^2$-minimization. 
Any larger number of components leads to a smoother description of the CPB spectrum but does not change the subsequent calculations.

The specific intensity of a blackbody is described by Planck's law,
\begin{equation}
    \nu I_\nu = \frac{2h\nu^4}{c^2} \frac{1}{\exp{(\frac{h\nu}{k_\mathrm{B}T})} - 1}\mathrm{,}
    \label{eq:Plancks_law}
\end{equation}
where $h$ is Planck's constant, $\nu$ is the frequency, $k_\mathrm{B}$ the Boltzmann constant, and $T$ is the temperature.
We introduce an amplitude $A$ as a free parameter, that scales the power of the blackbody functions.
This is typically referred to as a dilution factor, resulting in ``gray body spectra''.

Our full model reads
    \begin{equation}
    \nu I_\nu = \sum_{i=1}^{N_{\mathrm{BB}}} A_i \frac{2h\nu^4}{c^2} \frac{1}{\exp{(\frac{h\nu}{k_\mathrm{B}T_i})} - 1}\mathrm{,}
    \label{eq:full_spectral_model}
\end{equation}
and includes a total of 54 parameters, 27 temperatures and 27 amplitudes, and the fixed parameters $T_{\mathrm{CMB}}^0 = \SI{2.7255}{K}$ and $A_\mathrm{CMB}^0 = 1$ for the CMB.
This leads to a reasonable description of the overall spectral shape of the CPB without large residuals.
The full spectrum is shown in Fig.\,\ref{CPB redshift 0.0}.

\begin{table}[t]
\centering
\footnotesize
    \begin{tabular}{c|c|c|c|c}
        Nr. & Range & Sources & Composition & Refs. \\
        \hline \hline
        1 & \SI{37}{MHz} & Gal \& & 0.5\,Gal \& & (1) \\
        \cline{1-2}
        2 & \SI{295}{MHz} & AGN & 0.5\,AGN & (2)\\
        \hline
        3 & \SI{160}{GHz} & CMB & CMB & (3) \\
        \hline
        4 & \SI{264}{\upmu m} & Stars \& & \\
        \cline{1-2}
        5 & \SI{74.4}{\upmu m} & Gal, &  Gal & (4) \\
        \cline{1-2}
        6 & \SI{25.1}{\upmu m} & heated dust & \\
        \hline
        7 & \SI{4.09}{\upmu m} & Stars \& & \\
        \cline{1-2}
        8 & \SI{924}{nm} & Gal, &  Gal & (5)\\
        \cline{1-2}
        9 & \SI{271}{nm} & direct light &  \\
        \hline
        10 & \SI{57.9}{nm} & Gal/QSO & redshift-dependent & (6)\\
        \hline
        11 & \SI{13.1}{nm} & purely &  \\
        \cline{1-2}
        12 & \SI{3.22}{nm} & QSO & QSO & (6) \\
        \hline
        13 & \SI{767}{eV} & QSO, & 0.25\,Gal + 0.75\,QSO  \\
        \cline{1-2}
        14 & \SI{3.01}{keV} & gal. srcs. & 0.1\,Gal + 0.9\,QSO & (6,7) \\
        \hline
        15 & \SI{11.8}{keV} & purely & \\
        \cline{1-2}
        16 & \SI{39.6}{keV} & QSO & QSO & (7) \\
        \hline
        17 & \SI{134}{keV} & QSO \& & 0.1\,FSRQ + 0.9\,QSO \\
        \cline{1-2}
        18 & \SI{536}{keV} & FSRQ & 0.5\,FSRQ + 0.5\,QSO & (8) \\
        \hline
        19 & \SI{2.10}{MeV} & FSRQ \& & 0.25\,SNe + 0.75\,FSRQ \\
        \cline{1-2}
        20 & \SI{9.91}{MeV} & SNe Ia & 0.25\,SNe + 0.75\,FSRQ & (9) \\
        \hline
        21 & \SI{39.0}{MeV} &  & 0.67\,FSRQ + 0.33\,BL\,Lac  \\
        \cline{1-2}
        22 & \SI{132}{MeV} & FSRQ \& & 0.6\,FSRQ + 0.5\,BL\,Lac \\
        \cline{1-2}
        23 & \SI{481}{MeV} & BL\,Lac & 0.5\,FSRQ + 0.5\,BL\,Lac \\
        \cline{1-2}
        24 & \SI{1.81}{GeV} &  & 0.4\,FSRQ + 0.6\,BL\,Lac & (8,10)\\
        \hline
        25 & \SI{6.70}{GeV} & & \\
        \cline{1-2}
        26 & \SI{23.4}{GeV} & purely &  \\
        \cline{1-2}
        27 & \SI{71.8}{GeV} & BL\,Lac & BL\,Lac & (10) \\
        \cline{1-2}
        28 & \SI{236}{GeV} &  & \\
    \end{tabular}
    \caption{Parametrization of 28 blackbodies with main sources, corresponding composition and redshift evolution given in the following references: (1) \citep{Mancuso2015,Aversa2015}. (2) \citep{Yuan2017}. (3) \citep{Fixsen_2009}. (4) \citep{Gruppioni2013}. (5) \citep{Helgason2012}. (6) \citep{Khaire2019}. (7) \citep{Shen2020,Aird2015}. (8) \citep{Marcotulli2022,Ajello2009}. (9) \citep{Palicio2024,Ruiz-Lapuente2016}. (10) \citep{Ajello2014}.}
    \label{tab: Composition of BBs}
\end{table}

To eventually determine the blackbody evolution as a function of redshift, we calculate its mean energy by
\begin{equation}
    \langle E \rangle = \frac{\pi^4}{30 \zeta(3)} k_\mathrm{B} T \approx 2.701 \, k_\mathrm{B} T\mathrm{,}
    \label{eq:mean_energy_BB}
\end{equation}
which enables us to match each blackbody with their corresponding source composition in the local Universe.
We summarize our blackbody source-type matching in Tab.\,\ref{tab: Composition of BBs}.
We note that the uncertainties of the CPB spectrum directly impacts the amplitudes, and to a lesser extent the temperatures, of the 27 fitted components.
However, the temperature is a mere shape parameter and not necessarily physically meaningful (except for the CMB, CIB, and COB, for example). 
Therefore, the only uncertainty that may impact the final result are the amplitudes, which are typically uncertain by one order of magnitude, or more in the COB and CGB (MeV) range.

\subsection{CPB Redshift Evolution}\label{sec:redshift_evolution}
To model the isotropic CPB spectrum as a function of redshift, we need to include two different effects:
On the one hand, there is the effect of redshift itself, decreasing the photon energy as it propagates towards now.
The effect is modeled by stretching each blackbody by a factor of $(1+z)$. 
On the other hand, the sources whose light makes up the CPB at specific energy ranges, change in their abundance throughout the history of the Universe and, therefore, their luminosity changes as well.

For the latter, we investigate different luminosity functions of the various source types. 
In general, the luminosity function returns the number of sources per volume and per luminosity $L$.
A typical parametrization of a luminosity function is a  smooth double-power law,
\begin{equation}
    \phi(L) = \frac{\phi^*}{\left(\frac{L}{L^*}\right)^{\gamma_1} + \left(\frac{L}{L^*}\right)^{\gamma_2}}\mathrm{,}
    \label{eq:generic_luminosity_function}
\end{equation}
which is often used for AGNs, such as in QSO and blazars \citep{Ajello2009, Ajello2012, Ajello2014}.
The four parameters are the normalization $\phi^*$, the break-luminosity $L^*$, and the faint- and bright-end slopes $\gamma_1$ and $\gamma_2$.

Another often used parametrization is the Schechter function, which only uses the three parameters $\phi^*$, $L^*$, and $\alpha$.
It is defined as \citep{schechter1976analytic}
\begin{equation}
    \phi(L) = \phi^* \left(\frac{L}{L^*}\right)^\alpha \exp{\left(-\frac{L}{L^*}\right)}\mathrm{.}
    \label{sec:Schechter_function}
\end{equation}

With the integral
\begin{equation}
    L_\mathrm{total} = \int_{L_\mathrm{min}}^{L_\mathrm{max}} L \,\phi(L) \,\mathrm{d}L\mathrm{,}
    \label{eq:luminosity_integral}
\end{equation}
we are able to calculate the total luminosity $L_\mathrm{total}$ of a given source type.

There are several ways we can introduce an evolution parameter into the luminosity function. 
A simple parametrization is the pure density evolution (PDE), in which the luminosity function is multiplied by an evolution parameter, regardless of the different luminosities.
The PDE reads
\begin{equation}
    \phi_{\mathrm{PDE}}(L, z) = \phi(L, z = 0) \cdot e(z)\mathrm{.}
    \label{eq:PDE}
\end{equation}
A similar approach is the pure luminosity evolution (PLE), in which the luminosities themselves undergo an evolution, while the luminosity function remains unchanged,
\begin{equation}
    \phi_{\mathrm{PLE}}(L,z) = \phi(L/e(z), z = 0)\mathrm{.}
    \label{eqLPLE}
\end{equation}
A more sophisticated approach lies in the luminosity-dependent density evolution (LDDE).
This parametrization introduces an evolution parameter that also includes a dependency on the luminosity.
This is achieved by allowing each parameter of the evolution function to vary in some way as a function of luminosity,
\begin{equation}
    \phi_{\mathrm{LDDE}}(L,z) = \phi(L,z=0) \cdot e(L, z)\mathrm{.}
    \label{eq:LDDE}
\end{equation}
Additionally, there are approaches, that cannot be strictly categorised into one of these three categories. 
Alternatively, one could simply allow each of the parameters of the original luminosity function to vary in some way.

A deviation from this general approach is the redshift evolution of the CMB that can easily be modeled by multiplying the temperature of that blackbody by $(1+z)$, so that
\begin{equation}
    T_\mathrm{CMB}(z) = T_\mathrm{CMB,0} \cdot (1+z)\mathrm{,}
    \label{eq:CMB_evolution}
\end{equation}
where the temperature of the CMB for the local Universe is $T_\mathrm{CMB,0} = \SI{2.7255}{K}$ \citep{Fixsen_2009}.
This, by the nature of the blackbody spectrum, will also lead to an increased flux level, even though the multiplicative amplitude is fixed to $1$, while simultaneously shifting the spectrum to higher energies for the CMB.

\begin{figure}[h]
    \centering
    \includegraphics[width=9cm]{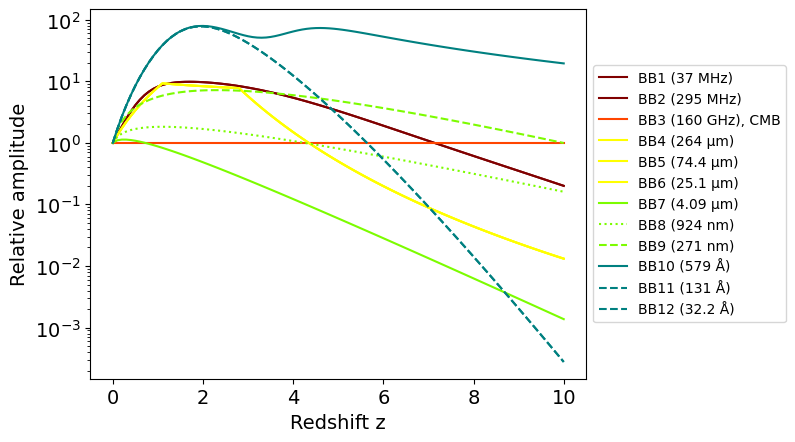}
    \includegraphics[width=9cm]{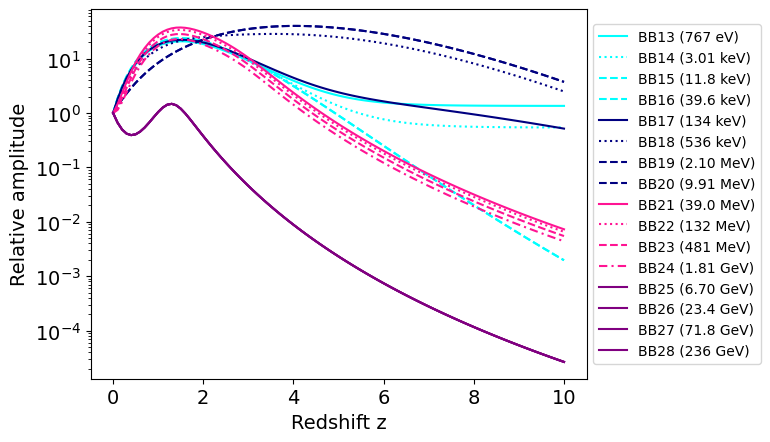}
    \caption{Evolution of amplitude relative to redshift 0 for the first 12 Blackbodies (top), and for Blackbodies 13 -- 28 (bottom). For most blackbodies, the evolution is loosely connected to the SFR density of the Universe, which shows an increase from present day up to $z \sim 2$ and a decrease afterwards. A big exception is Blackbody 3 (orange), the component modelling the evolution of the CMB, which is entirely described by an increase with temperature, while keeping the amplitude fixed.}
    \label{fig: Amplitudes}
\end{figure}

\begin{figure}
    \centering
    \includegraphics[width=8.5cm]{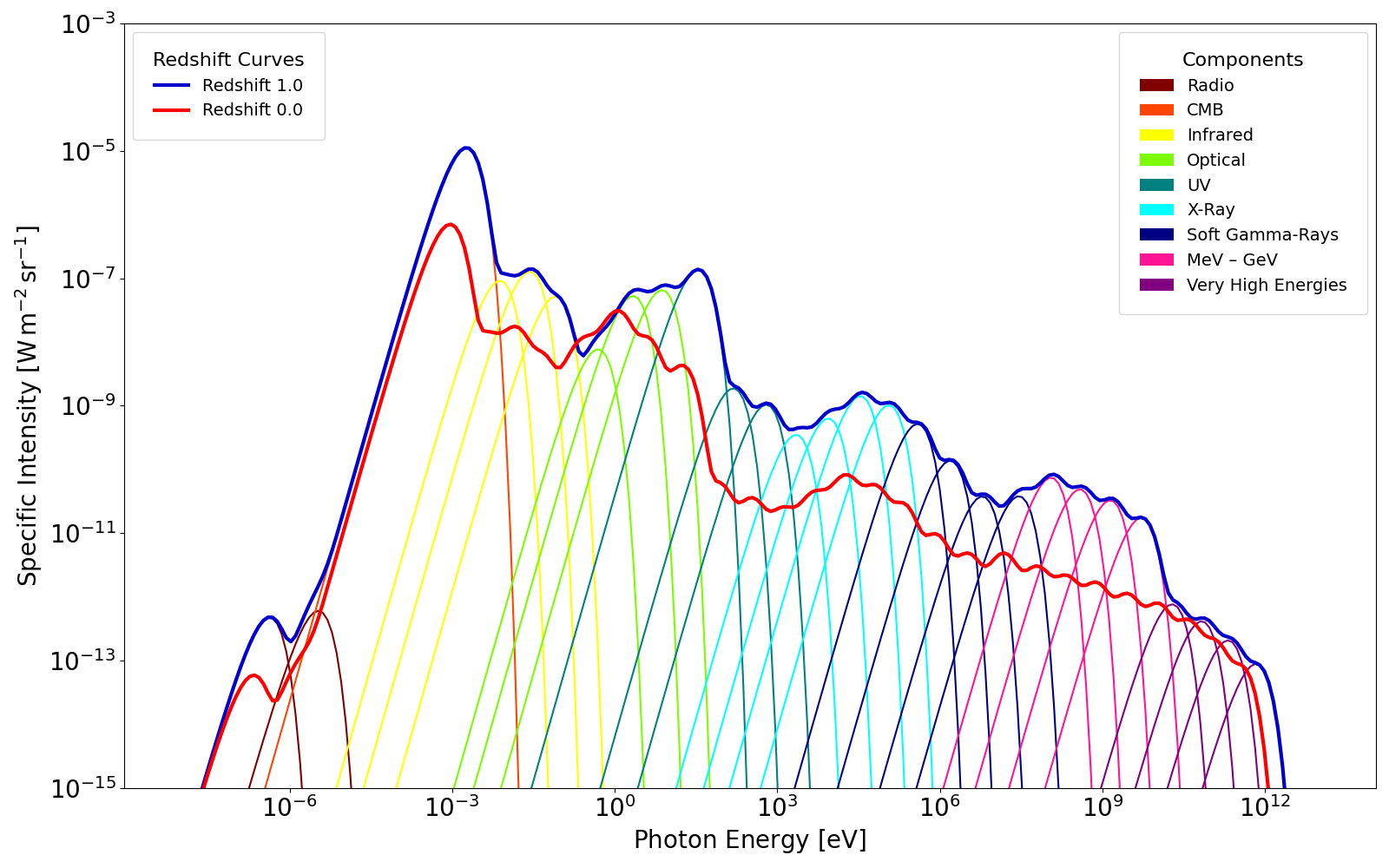}
    \includegraphics[width=8.5cm]{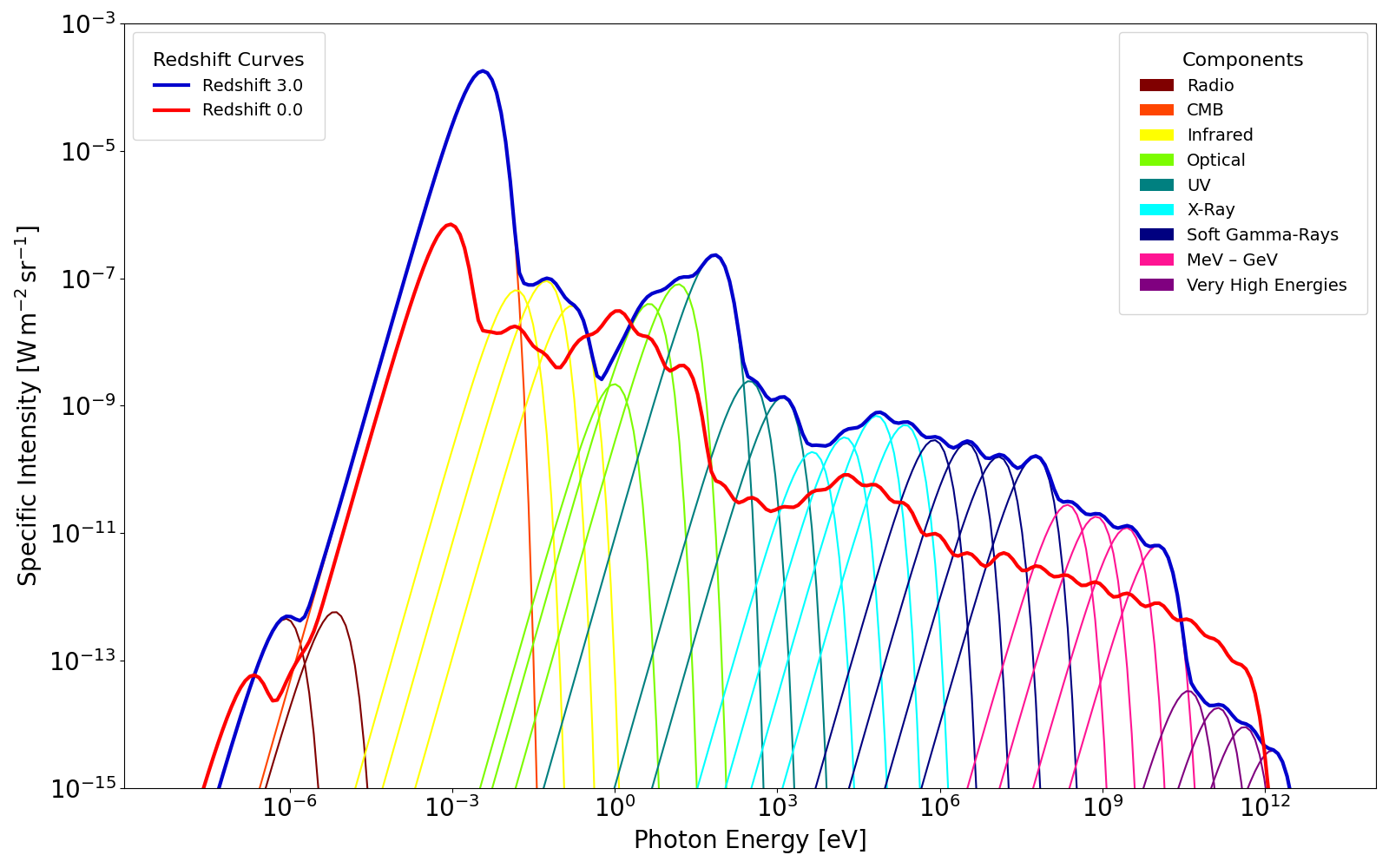}
    \includegraphics[width=8.5cm]{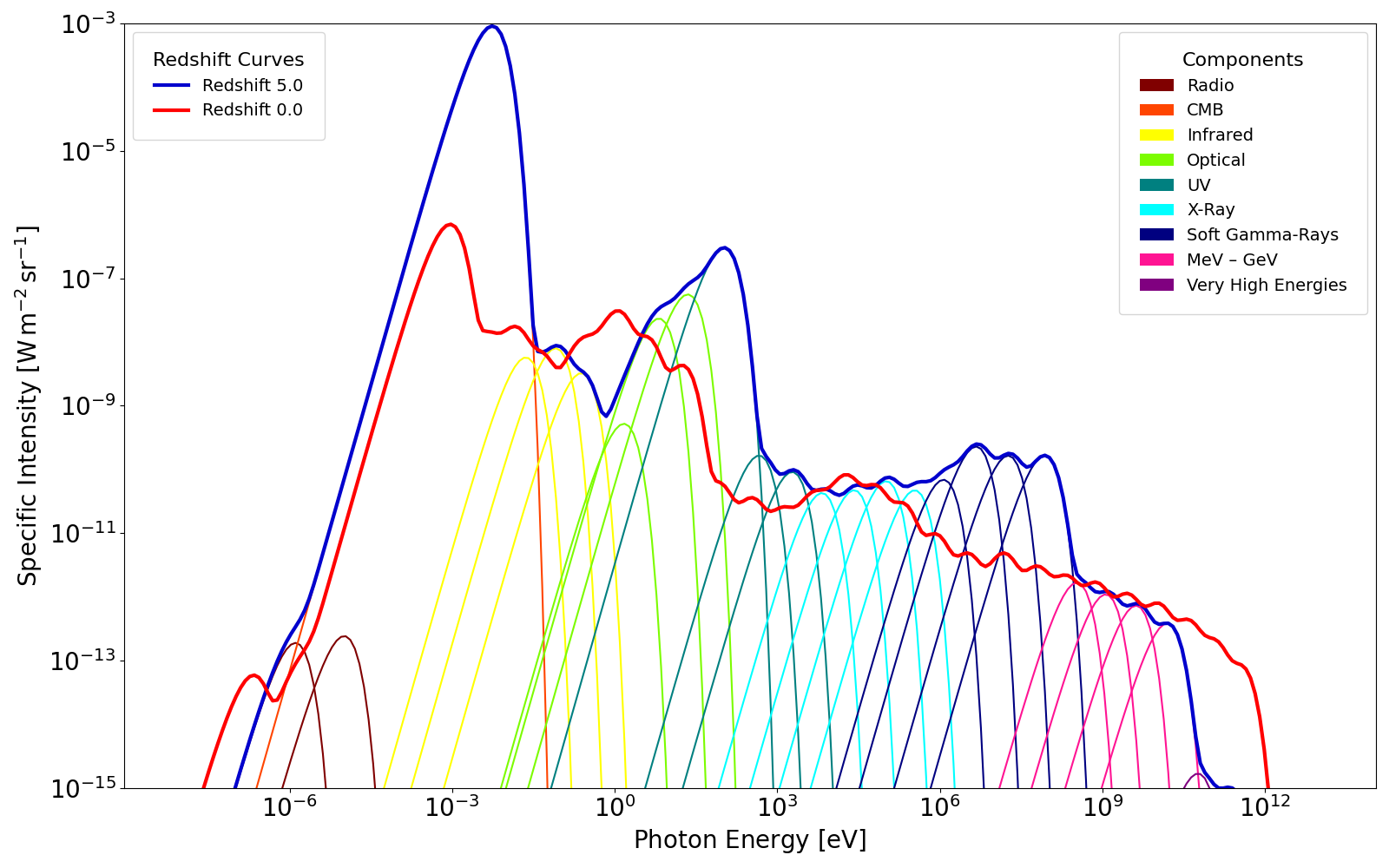}
    \includegraphics[width=8.5cm]{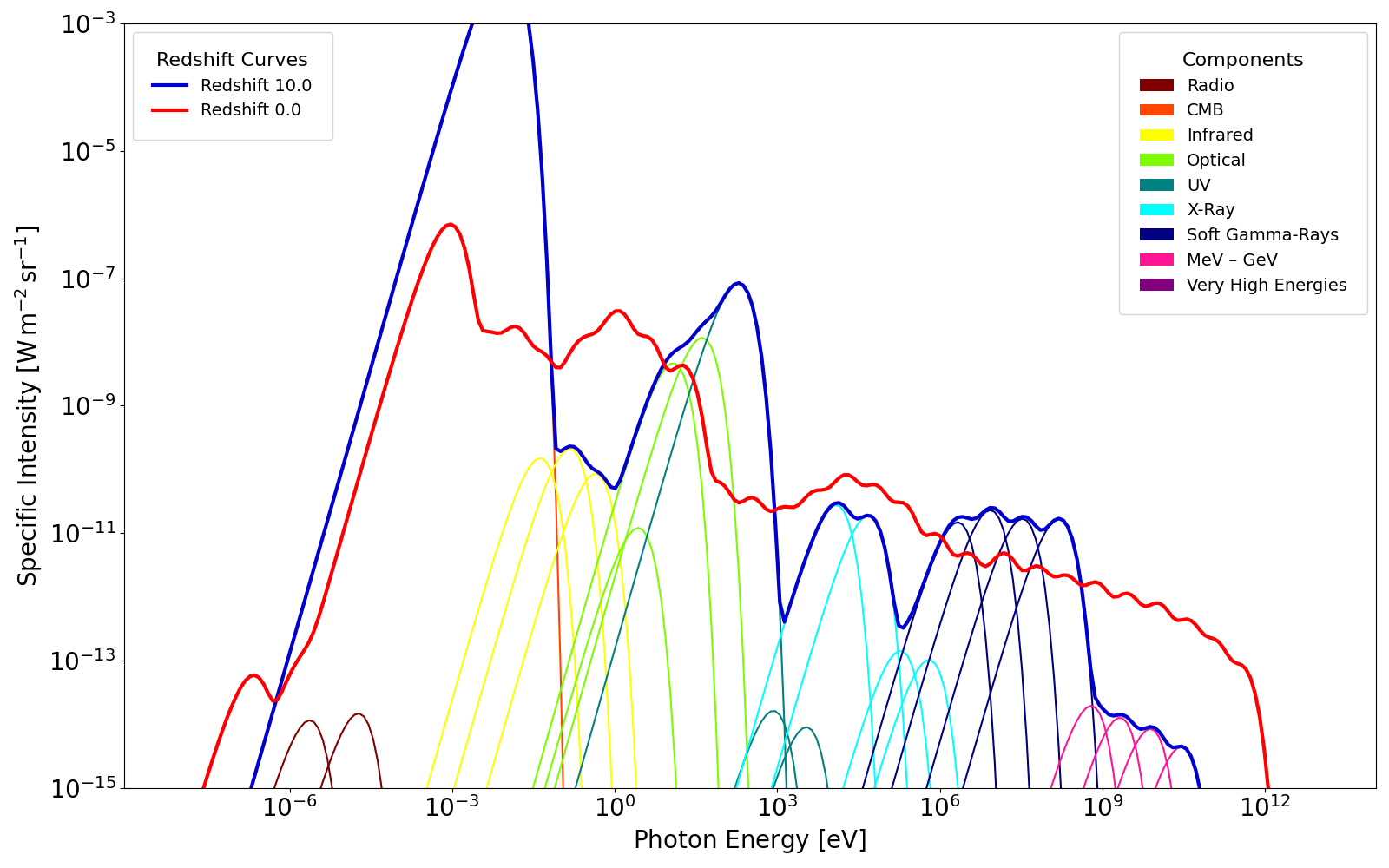}
    \caption{Redshift evolution for the CPB for $z=1$, $z=3$, $z=5$, and $z=10$ (blue) in comparison to $z=0$ (red). Up to $z \sim 3$, most blackbody components have reached peak amplitude. At $z=5$, most amplitudes are comparable to $z=0$, while at $z=10$ most components decline again, except for some UV and soft gamma-ray components that remain elevated. In contrast, there is the behavior of the CMB, which continues to increase in intensity and shift to higher energies with redshift.}
    \label{fig: four redshifts CPB}
\end{figure}

We calculate the luminosities for redshifts up to $z = 10$ in $\Delta z = 0.1$-steps for different sources and energies.
This allows us to determine the relative change of the amplitudes for each component according to our model, which we summarize in Tab.\,\ref{tab: Composition of BBs}.
The detailed functional forms used are given in respective references.
The relative amplitude $A(z)/A(z=0)$, i.e. the change for each blackbody relative to redshift $z=0$, is shown in Fig.\,\ref{fig: Amplitudes}.
Combining these amplitudes and the redshift evolution, we are able to construct the CPB for up to $z = 10$, some of which we show in Fig.\,\ref{fig: four redshifts CPB}.
We note that beyond $z=5$, the luminosity functions become rather uncertain.
According to our decomposition, at $z = 10$, one would expect the CMB to completely dominate the spectrum, together with a peak in the UV as well as around MeV (see Fig.\,\ref{fig: four redshifts CPB}).

Naturally, a striking connection can be made to the Madau-plot \citep{Madau2014}, which shows the SFR density of the Universe as a function of redshift.
Starting from $z = 0$, the SFR increases up to a redshift of about 2, after which it decreases.
Since the luminosity evolution of most objects in the Universe is linked to star formation, the CPB will therefore also show a similar behavior.
Likewise, the differential pair production rate as a function of redshift (see Fig.\,\ref{fig:emissivity_redshift}) should also clearly follow the SFR density.

\subsection{Integrating over the Co-moving Volume}\label{sec:comoving_volume}
To determine the positron production rate in a given volume, we integrate over the corresponding co-moving volume of space.
For that, we assume the emissivity $\dot{n}$ to be constant in the interval $\Delta z = \pm 0.05$ around a given redshift.
We compute the co-moving volume in a given redshift bin by integrating the differential co-moving volume \citep{hogg1999distance},
\begin{equation}
    \frac{\mathrm{d}V_\mathrm{com}}{\mathrm{d}z \, \mathrm{d}\Omega} = \frac{c \cdot D_\mathrm{com}^2(z)}{H(z)}\mathrm{,}
    \label{eq:comoving_Jacobian}
\end{equation} 
where $D_\mathrm{com}(z)$ is the co-moving distance, defined as
\begin{equation}
    D_\mathrm{com}(z) = \int_0^z \frac{c}{H(z')} \mathrm{d}z'\mathrm{.}
    \label{eq:comoving_distance}
\end{equation}
Here, $H(z)$ is the Hubble parameter,
\begin{equation}
    H(z) = H_0 \sqrt{\Omega_m(1+z)^3 +\Omega_\Lambda}\mathrm{,}
    \label{eq:hubble_parameter}
\end{equation}
with $H_0 = \SI{67.66}{\frac{km/s}{Mpc}}$, $\Omega_m = 0.31$, and $\Omega_\Lambda = 0.69$ \citep{Planck2020} for a standard $\Lambda$CDM-Universe.
%
%Therefore, the co-moving volume $V_\mathrm{com}$ can be calculated by integrating over the differential co-moving volume,
%
%\begin{equation}
%    V_\mathrm{com}(z) = 4\pi\int_{z_\mathrm{center}-\Delta z}^{z_\mathrm{center}+\Delta z} \frac{\mathrm{d}V_\mathrm{com}}{\mathrm{d}z \, \mathrm{d}\Omega}\,\mathrm{d}z\mathrm{.}
%\end{equation}

\begin{figure}[]
    \centering
    \includegraphics[width=1.0\linewidth]{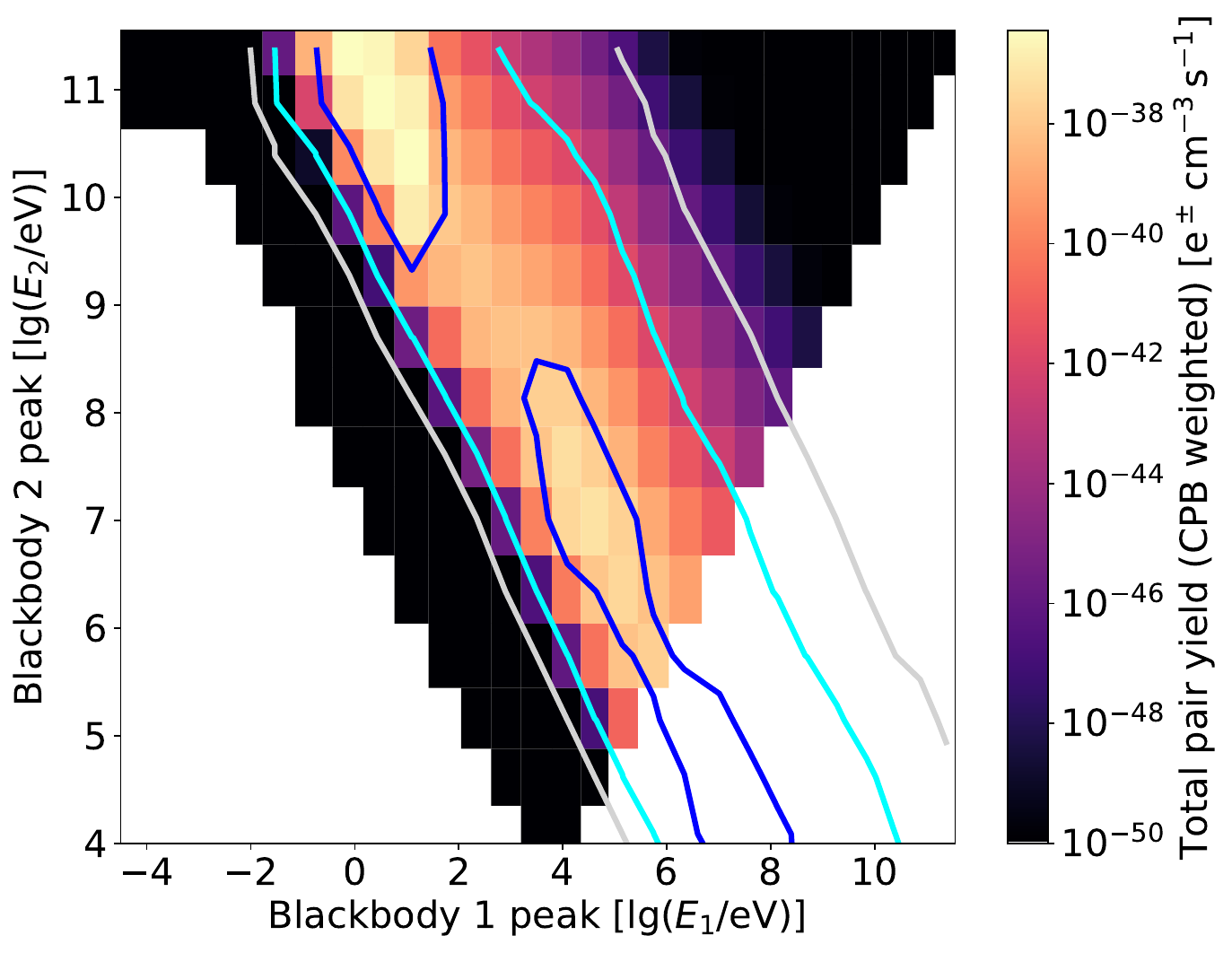}
    \caption{Pair yield contribution from diluted blackbodies interacting among themselves as a function of blackbody temperature for $z=0$. The levels are $10^{-38}$ (blue), $10^{-42}$ (cyan), and $10^{-46}\,\mathrm{e^\pm\,cm^{-3}\,s^{-1}}$ (gray), respectively. The two maxima indicate the high-energy bump above 1\,GeV (top left) and the low-energy, power-law-index $-1$ tail (bottom right). The peak temperature of the blackbodies is calculated via $E_{1/2} \approx 2.82\,k_\mathrm{B}T_{1/2}$ for illustration purpose.}
    \label{fig:pair_yield_contribution}
\end{figure}

\subsection{Pair Injection Spectrum}\label{sec:pair_injection_spectrum}
\begin{figure}
    \centering
    \includegraphics[width=1.0\linewidth]{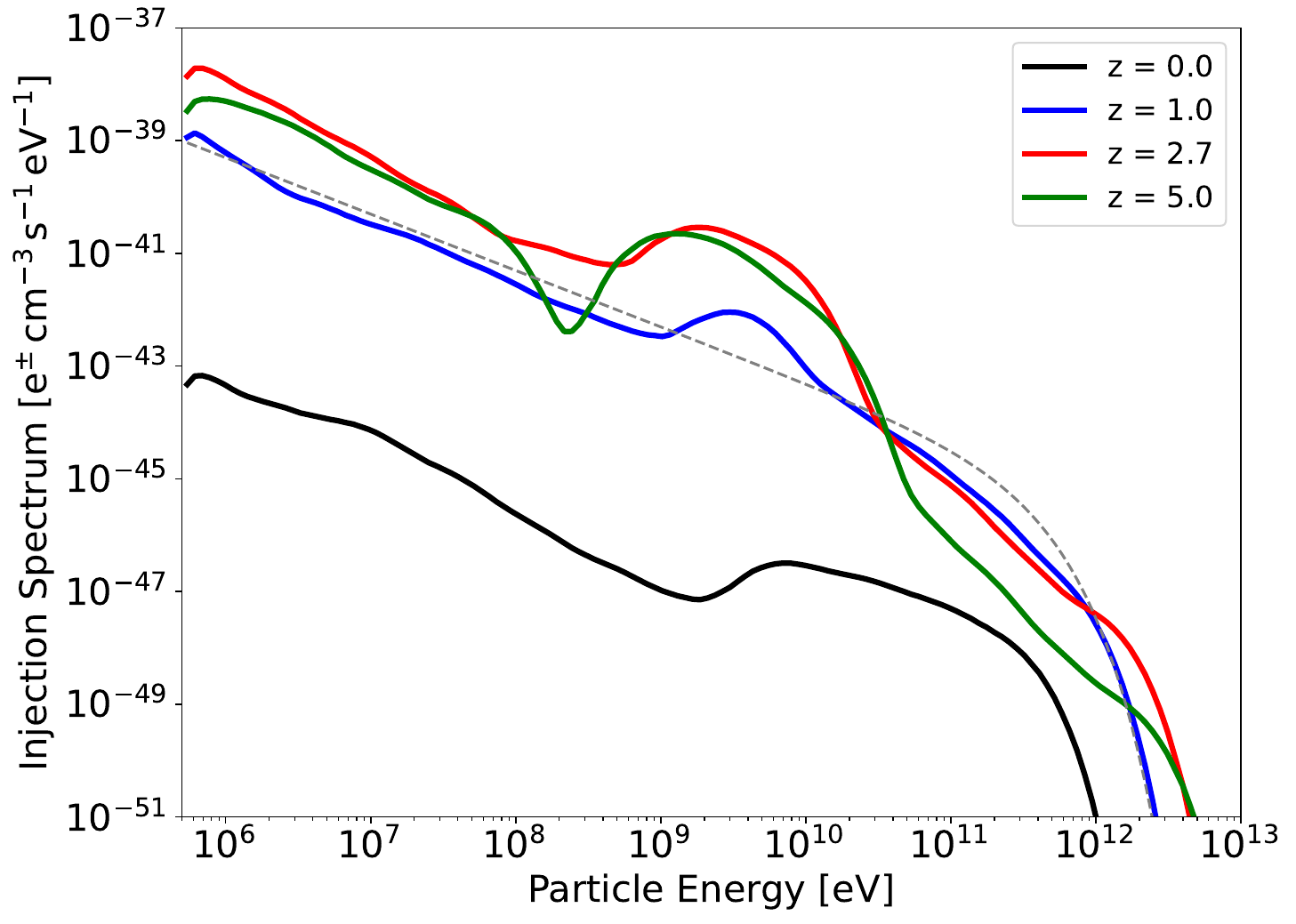}
    \caption{Pair injection spectrum from $\gamma$-$\gamma$ pair-production of the CPB with itself. The general shape (dashed gray line) can be described by a cutoff power-law with index $-1$ and cutoff energy around $0.1$--$1$\,TeV.}
    \label{fig:pair_injection_spectrum}
\end{figure}

We now use Eq.\,(\ref{eq:final_spectrum_integral}) for each redshift bin to calculate the cosmological pair injection spectrum from $\gamma$-$\gamma$ pair-production.
We show the relative contribution of all possible intercombinations of black bodies interacting among themselves for $z=0$ in Fig.\,\ref{fig:pair_yield_contribution}.
It becomes clear that the highest contribution comes from temperatures around $10^4$\,K and $10^{14}$\,K, corresponding to energies around $2$--$3$\,eV and $20$--$30$\,GeV, respectively.
This is not a contradiction to the pair creation threshold, as the blackbody spectra extend to higher energies, and is a mere reflection of the choice of the decomposition into spectral components.
The injection spectrum for different redshifts is shown in Fig.\,\ref{fig:pair_injection_spectrum}.
The local pair spectrum is about four orders of magnitude weaker than at higher redshifts.
The peak is found around redshift $z = 2.7$.
The general shape is a cutoff power-law with index $-1$ and cutoff energies around $0.1$\,TeV at redshift $z = 0$, up to $1$\,TeV above redshift $z = 5.0$.
However, the spectrum is not generally smooth because of the two main contributing regions from the CPB:
While the high-energy bump at $z = 0$ above 1\,GeV is made of the entire CGB interacting with the CIB and COB, the low-energy part originates from the cumulative effect of all remaining interactions.
The same is true for the other redshifts at different scalings.
The high-energy bumps make a sizable contribution to the overall intergalactic cosmic-ray electron and positron spectrum as more than 50--70\% of the pairs are found above 1\,GeV.

\begin{figure}[!hb]
    \centering
    \includegraphics[width=1.0\linewidth]{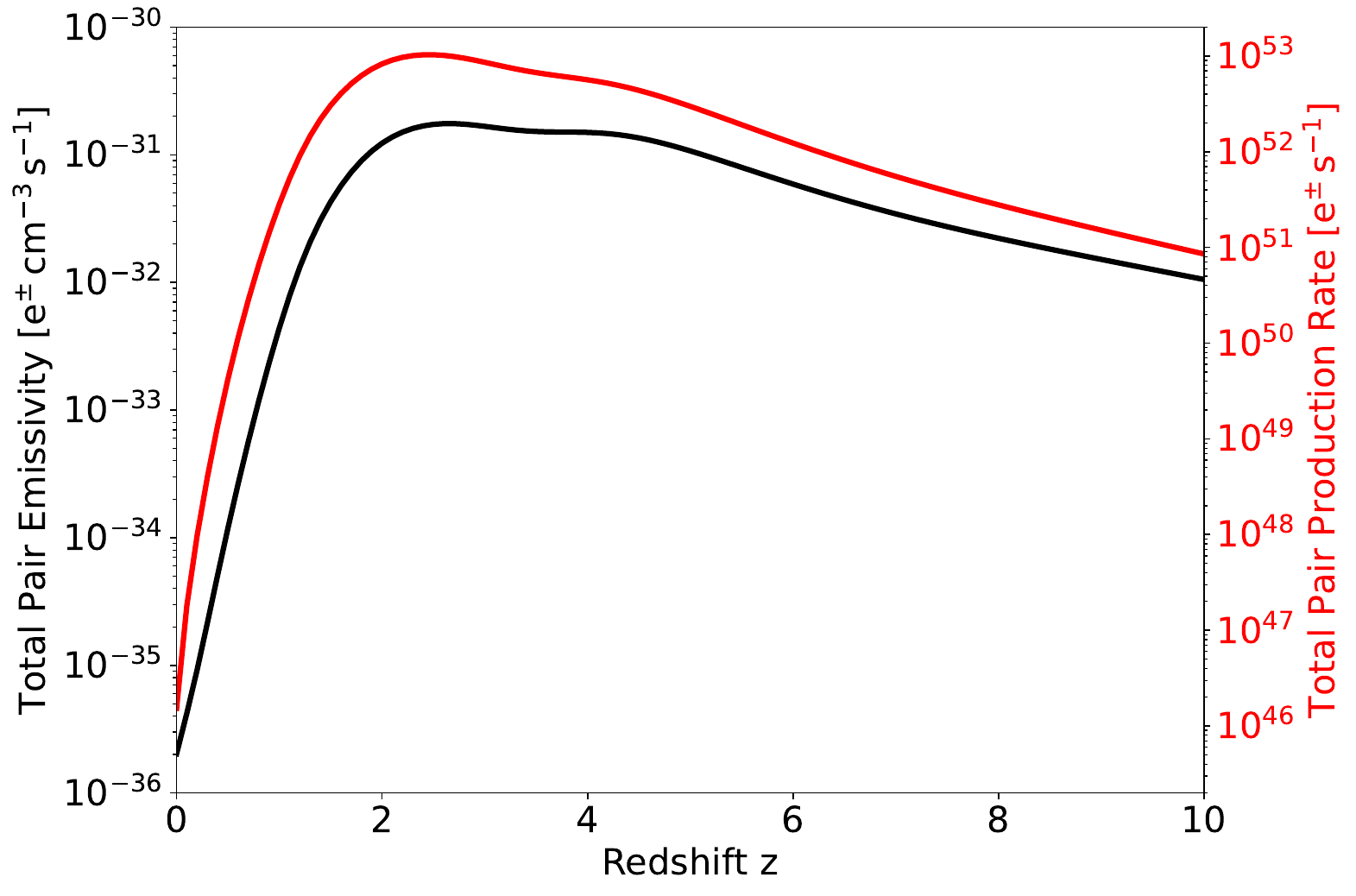}
    \caption{Total pair emissivity (left axis, black) and total pair production rate (right axis, red) from $\gamma$-$\gamma$ pair-production as a function of redshift per redshift bin of $\Delta z = 0.1$. The local contribution therefore integrates up to a distance of $\approx 200$\,Mpc.}
    \label{fig:emissivity_redshift}
\end{figure}

The total local pair emissivity is $2 \times 10^{-36}\,\mathrm{cm^{-3}\,s^{-1}}$.
Then, the emissivity shows a steep increase for up to a redshift of $2.7$, after which a slight decrease can be observed again.
We show the total pair emissivity as a function of redshift in Fig.\,\ref{fig:emissivity_redshift}, together with the corresponding total pair production rate.
The local pair production rate up to a distance of $\approx 200$\,Mpc ($z < 0.05$) is then on the order of $10^{46}\,\mathrm{e^\pm}\,s^{-1}$.
Scaling this volumetrically to the Milky Way up to $50$\,kpc results in a possible Galactic contribution on the order of $10^{35}\,\mathrm{e^\pm\,s^{-1}}$.
This is about eight orders of magnitude smaller than the Milky Way annihilation rate \citep{Siegert2016_511}.
We can therefore safely conclude that this cosmological process hardly contributes to the local \emph{positron puzzle}.
However, since the pair production rate is about seven orders of magnitude stronger around redshifts $z = 2$--$5$, the secondary (Inverse Compton and bremsstrahlung) and tertiary (annihilation in flight and at rest) emission of those particles may contribute significantly to the CGB.
We compute the expected emission signatures in the following.

\section{Secondary Emission}\label{sec:secondaries}
\begin{figure}[!b]
    \centering
    \includegraphics[width=1.0\linewidth]{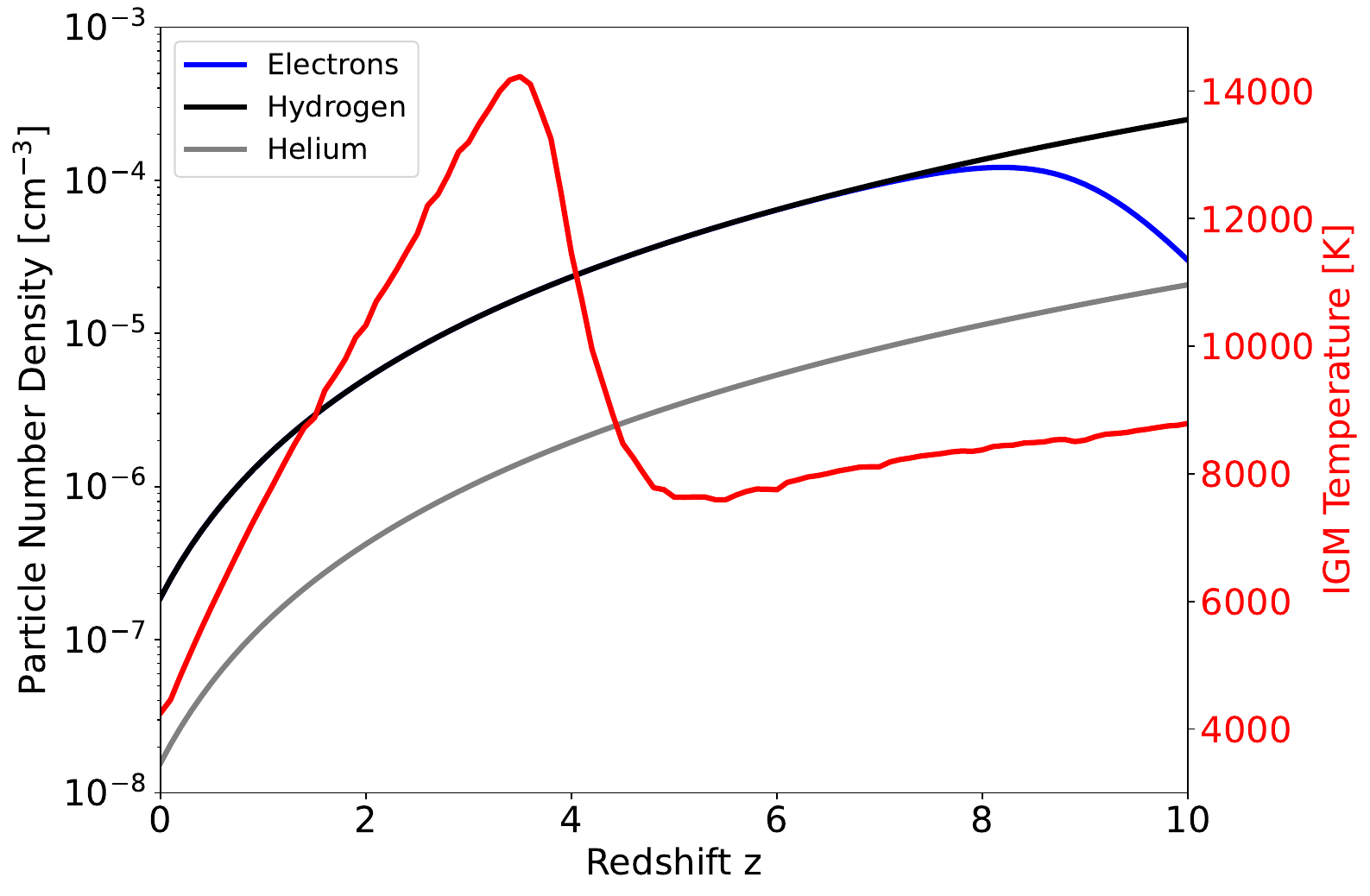}
    \caption{Intergalactic medium minimum particle densities according to $\Lambda$CDM (left axis) for electrons, hydrogen and helium, and temperature (right axis, red) as a function of redshift. The era of re-ionization is visible by the electron density.}
    \label{fig:IGM_models}
\end{figure}

We solve the homogeneous isotropic cosmological continuity equation,
\begin{eqnarray}
    \frac{\partial N(E,z)}{\partial t} &+& \frac{\partial}{\partial E}\left(\dot{E}(E,z)\,N(E,z)\right) + 3H(z)\,N(E,z) = \\
    & = & Q(E,z) - \Gamma_\mathrm{ann}(E,z)\,N(E,z)\mathrm{,}
    \label{eq:cosmo_continuity_equation}
\end{eqnarray}
in proper variables and switch to co-moving differential energy density $Y(E,z) \equiv a^3(z)\,N(E,z)$ to arrive at
\begin{equation}
    \frac{\partial Y(E,z)}{\partial t} + \frac{\partial}{\partial E}\left(\dot{E}(E,z)\,Y(E,z)\right) = a^3(z)\,Q(E,z) - \Gamma_\mathrm{ann}(E,z)\,Y(E,z)\mathrm{.}
    \label{eq:conti_eq_reduced}
\end{equation}
In Eq.\,(\ref{eq:cosmo_continuity_equation}), $N(E,z) \equiv \mathrm{d}n_{e^+}/\mathrm{d}E$ is the isotropic differential positron/electron number density in units of $\mathrm{e^\pm\,cm^{-3}\,eV^{-1}}$.
The injection function $Q(E,z)$ in units of $\mathrm{e^\pm\,cm^{-3}\,s^{-1}\,eV^{-1}}$ has been derived in the previous section, and shown in Fig.\,\ref{fig:pair_injection_spectrum}.
The sink term in the equations corresponds to the positron annihilation rate,
\begin{equation}
    \Gamma_\mathrm{ann}(E,z) = n_{e^-}(z)\,v(E)\,\sigma_\mathrm{ann}(E)\mathrm{,}
    \label{eq:pos_ann_rate}
\end{equation}
where $\sigma_\mathrm{ann}(E)$ is the cross section for direct annihilation with free electrons of density $n_{e^-}(z)$.
Thermal positronium (Ps) formation is applied to a separate ``thermal pool'' during the propagation of Eq.\,(\ref{eq:conti_eq_reduced}) if positrons are transferred below a threshold energy.
The thermal pool is depleted with a rate $\Gamma_{\mathrm{Ps,th}}(z,T(z),x(z))$ that depends on the IGM temperature $T(z)$ and ionization state $x(z)$.
We use the models by \citet{Puchwein2015_IGM} and \citet{Silva2013_IGM} to characterise $T(z)$ and $x(z)$, respectively, and use the coefficients for Ps formation via radiative recombination (rr) with free electrons, and via charge exchange with natural hydrogen (cxh) and helium (cxhe) from \citet{Wallyn1996_Ps} to compute $\Gamma_{\mathrm{Ps,th}}(z,T(z),x(z))$.
The IGM models for temperature and density are shown in Fig.\,\ref{fig:IGM_models}.
The thermal pool follows the differential equation
\begin{equation}
    \frac{\mathrm{d}n_\mathrm{th}}{\mathrm{d}t} = -\Gamma_{\mathrm{Ps,th}}(z)\,n_\mathrm{th}\mathrm{,}
    \label{eq:thermal_ode}
\end{equation}
so that it is decreasing exponentially as 
\begin{equation}
    n_\mathrm{th}(t(z)) = n_\mathrm{th,0}\,\exp\left[-\Gamma_\mathrm{Ps,th}(z)\,t(z) \right]
\end{equation}
and replenished by cooling below the threshold.
The lost thermal positrons will form Ps so that there will be an ortho-Ps and para-Ps spectrum \citep{Ore1949_Ps} as a function of redshift.

The continuous energy losses \citep[see, e.g.,][for summaries on positron energy losses]{Jean2009, Prantzos2011_511,Siegert2023},
\begin{equation}
    \dot{E}(E,z) = b(E,z) \equiv -\frac{\mathrm{d}E(E,z)}{\mathrm{d}t}\mathrm{,}
    \label{eq:energy_loss_def}
\end{equation}
consider ionization and plasma losses, bremsstrahlung in partly or fully ionized gas, synchrotron radiation in the intergalactic magnetic field, Inverse Compton scattering off the same CPB field, and adiabatic cooling by the Hubble flow.
The latter is generally described by
\begin{equation}
    b_\mathrm{AD}(E,z) = H(z)\,p(E)\,\frac{\partial E}{\partial p}\mathrm{,}
    \label{eq:adiabatic_cooling}
\end{equation}
where $H(z)$ is again the Hubble parameter from Eq.\,(\ref{eq:hubble_parameter}), and $p$ is the particle momentum.

\begin{figure}
    \centering
    \includegraphics[width=1.0\linewidth]{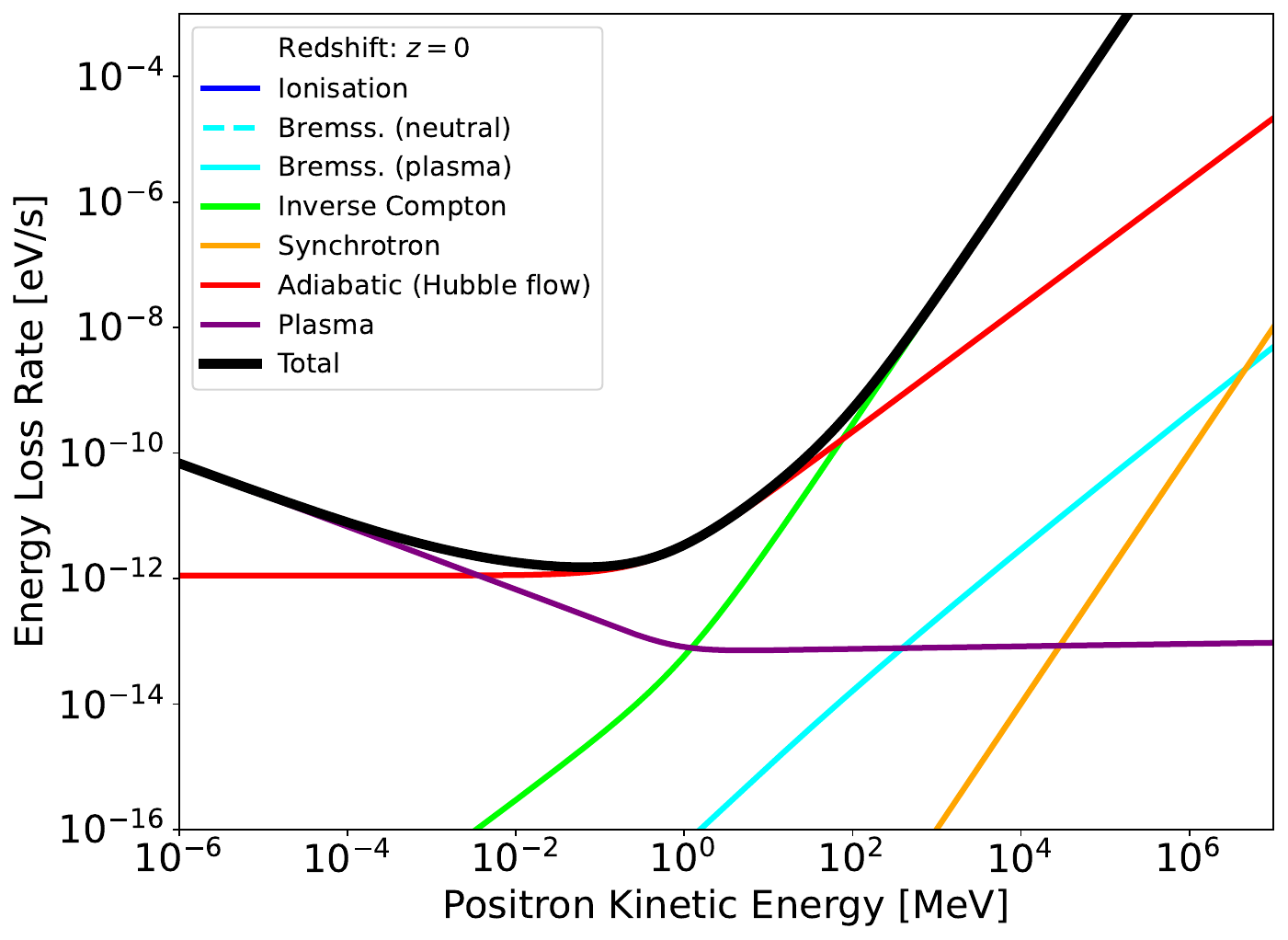}\\
    \includegraphics[width=1.0\linewidth]{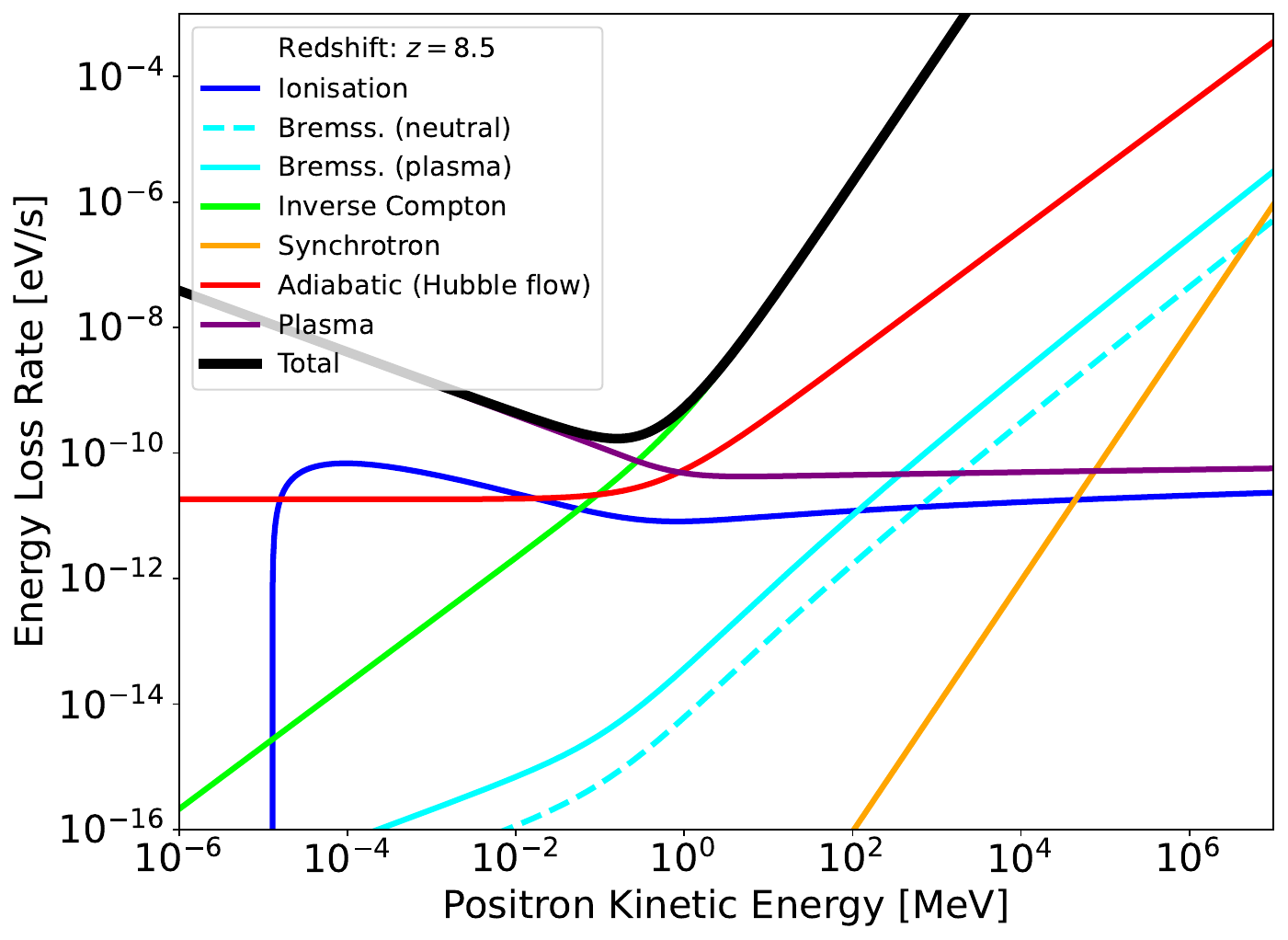}\\
    \caption{Cosmological energy loss rates for different processes at redshift $z=0$ (top) and $z=8.5$ (bottom).}
    \label{fig:cooling_functions}
\end{figure}

The energy losses depend on the cosmological model chosen, and in particular the IGM temperature, ionization state, magnetic field strength, and radiation fields.
We show the resulting cooling functions exemplarily for redshift $z = 0$ and $z = 8.5$ in Fig.\,\ref{fig:cooling_functions} to highlight the relative importance at different cosmological eras, that is before and after re-ionization.
It is evident that the minimum cooling rate is typically found around MeV energies, independent of redshift.
At low redshifts and the local Universe, the dominant cooling processes are plasma losses below $10^{-2}$\,eV kinetic energy, the Hubble flow up to $\sim 100$\,MeV, and Inverse Compton above.
At high redshifts beyond $z \approx 5$--$6$, ionization losses in partly ionized material also adds to the low-energy losses, however compared to plasma losses, this is still sub-dominant.
In general, the magnitude of the losses increases with redshift.
Due to the low number densities of baryons and electrons as the lowest limit of particle interactions, bremsstrahlung is never a strong cooling mechanism in the IGM.
Likewise, synchrotron losses are several orders of magnitude weaker than Inverse Compton losses as the mean IGM magnetic field is suspected to be around $10^{-15}$\,G \citep{Neronov2010_IGM_Bfield}.
However, the IGM magnetic field could also be on the order of $10^{-9}$\,G \citep{Locatelli2021_IGM_Bfield}, which would make synchrotron as highly cooling as Inverse Compton and lead to a strong radio background.
In this study, we assume negligible synchrotron cooling.
Since the main low-energy cooling mechanisms only depend on temperature, and the IGM temperature only varies within a factor of three to four in almost all models (Fig.\,\ref{fig:IGM_models}), there will be hardly any systematic uncertainty for lower energies.
Likewise, the high-energy cooling only depends on the radiation fields, of which the CMB is the strongest intergalactic component at all redshifts.
Therefore, the general shape and amplitude of the cosmological cooling function is rather robust.

\begin{figure*}[!ht]
    \centering
    \includegraphics[width=1\linewidth]{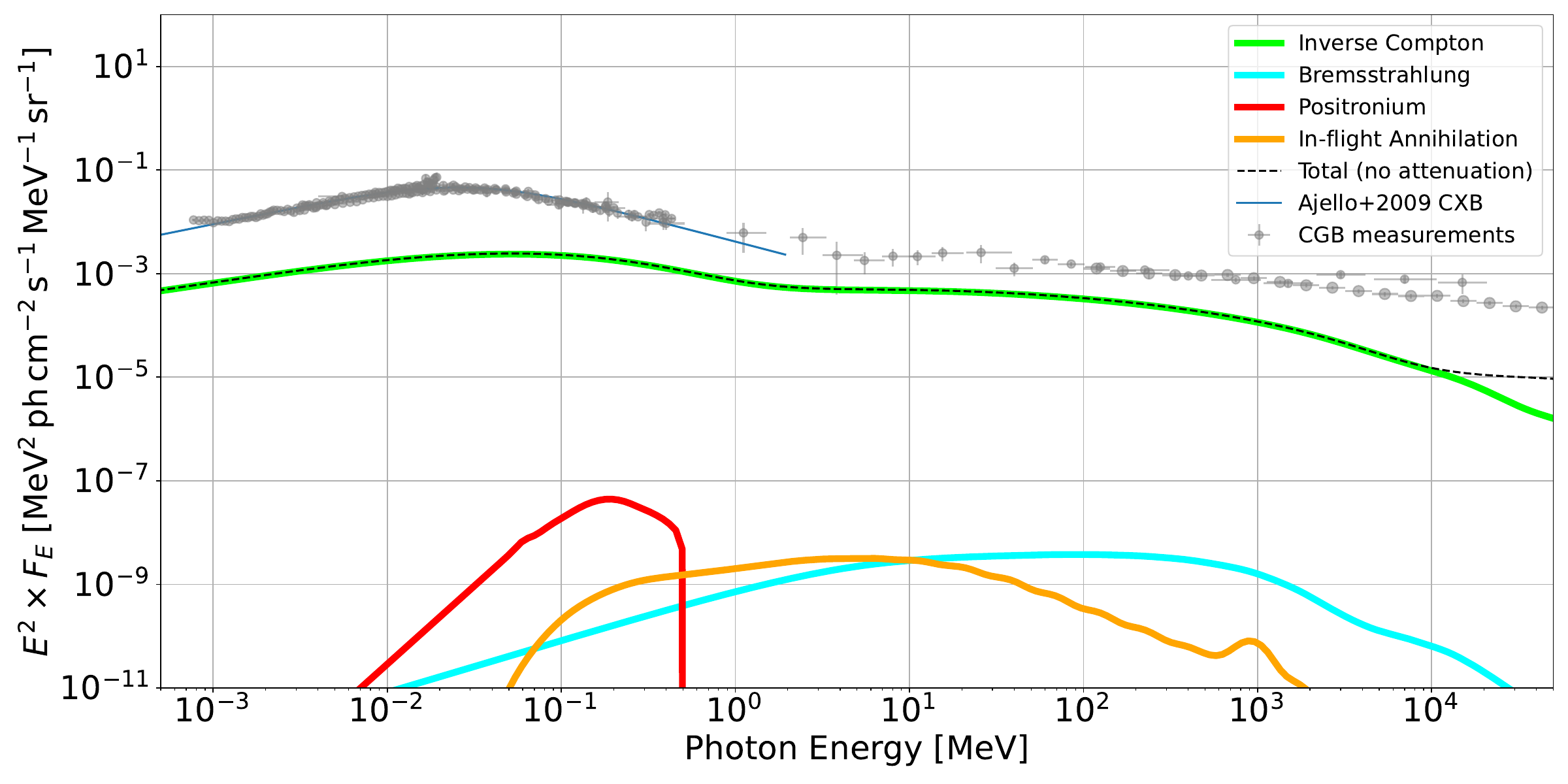}
    \caption{Secondary emission from pairs produced via $\gamma$-$\gamma$ absorption of the CPB with itself. The total, unabsorbed, spectrum is shown as dashed black line. The emission is dominated by Inverse Compton scattering off the CPB fields (green), with minor contributions from positron annihilation (red and orange) and bremsstrahlung (cyan). Previous CGB measurements are shown as gray data points from a compilation by \citet{Ajello2009} {and \citet{Ackermann_2015}}.}
    \label{fig:final_spectrum}
\end{figure*}

\begin{figure*}[!]
    \centering
    \includegraphics[width=1\linewidth]{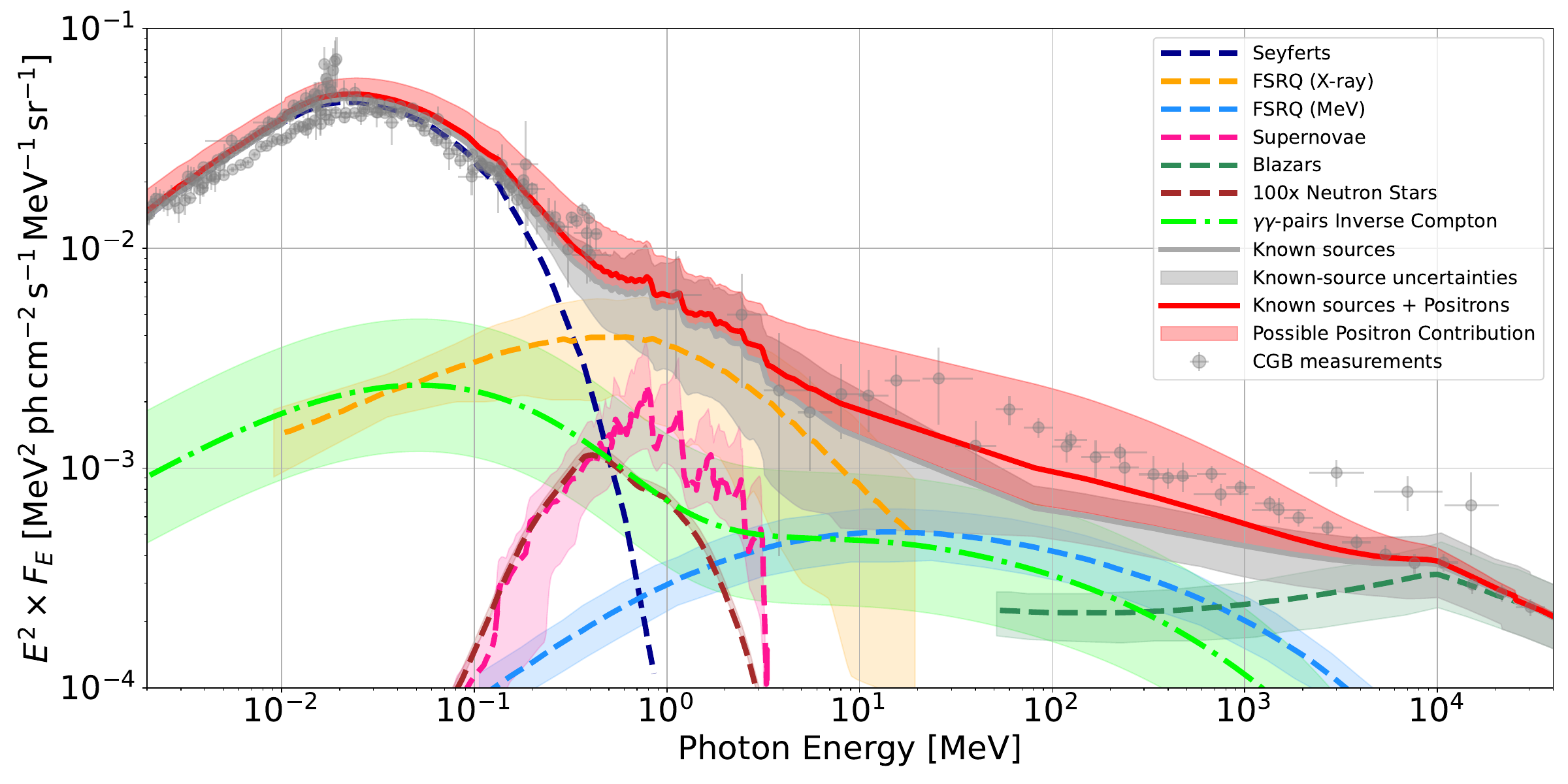}
    \caption{{A composition of the CPB from $\SI{20}{keV}$ to $\SI{40}{GeV}$. The individual contributions from established astrophysical source populations are shown separately with dashed lines, including Seyfert galaxies (darkblue) from \citet{Gilli+2007}, X-ray FSRQs (orange) from \citet{Ajello2009}, FSRQs (blue) from \citet{Ajello2012}, supernovae (pink) from \citet{Ruiz-Lapuente2016}, blazars (darkgreen) from \citet{Qu+2019}, and neutron stars (brown) from \citet{Ruiz-Lapuente_2020}. Shaded regions indicate uncertainties from the respective paper. The combined intensity is represented by a gray line. The emission from Inverse Compton scattering, which is the dominant energy loss process of positrons, is shown as a green dash-dotted line. The sum of the known-source background and the IC contribution defines the total expected emission, shown as the red band. The red shaded region indicates the uncertainty range associated with the IC component, corresponding to 0.075–5.1 times the fiducial IC contribution added to the known-source background. Measurements of the cosmic gamma-ray background from multiple instruments are overlaid as data points with corresponding error bars from \citet{Ajello2009} with some additional data points from \citet{Ackermann_2015}.}}
    \label{fig:MeVgap}
\end{figure*}

The next step is to calculate the emissivities for Inverse Compton, bremsstrahlung, and in-flight annihilation.
The former two are calculated by
\begin{equation}
    \epsilon_\mathrm{X}(E_\gamma,z) = \int\,\mathrm{d}E\,N(E,z)\,K_\mathrm{X}(E,E_\gamma,z)\mathrm{,}
    \label{eq:ic_br_syn_emiss}
\end{equation}
with $X = \left\{ \mathrm{IC,BR}\right\}$, and $K_\mathrm{X}(E,E_\gamma,z)$ being the corresponding integration kernel.
We use the delta-approximation \citep{Dermer1993_delta_IC} for the Inverse Compton emissivity and an approximate bremsstrahlung kernel that depends on the energy loss and is proportional to $E_\gamma^{-1}$.
We omit synchrotron radiation here as the contribution is expected to be small.
For in-flight annihilation, we use the full differential cross section to calculate 
\begin{equation}
    \epsilon_{\mathrm{IA}}(E_\gamma,z) = \int\,\mathrm{d}E\,N(E,z)\,n_{e^-}(z)\,v(E)\,\frac{\mathrm{d}\sigma_\mathrm{ann}(E,E_\gamma)}{\mathrm{d}E_\gamma}\mathrm{.}
    \label{eq:IA_emissivity}
\end{equation}

Finally, the differential emissivities as a function of redshift are integrated over the cosmology to obtain the isotropic differential flux via
\begin{equation}
    \frac{\mathrm{d}F(E_0)}{\mathrm{d}E_0} = \frac{c}{4\pi}\int_0^{z_\mathrm{max}}\,\mathrm{d}z\,\frac{\epsilon\left(E_\gamma = (1+z)E_0,z\right)}{H(z)(1+z)}e^{-\tau(E_0,z)}\mathrm{.}
    \label{eq:isotropic_flux}
\end{equation}
In Eq.\,(\ref{eq:isotropic_flux}), $E_0$ is the observed photon energy and $\tau(E_0,z)$ is the opacity of the Universe to $\gamma$-$\gamma$ pair-production (see Sect.\,\ref{Background}).
We show the final spectrum in the range between 0.5\,keV and 50\,GeV in Fig.\,\ref{fig:final_spectrum}.

{
In the keV-MeV energy range many different sources contribute significantly to the CPB.
These individual components are Seyferts, X-Ray FSRQs, FSRQs, supernovae, blazars, and neutron stars. 
Yet, their combined intensity cannot fully explain the CGB measurements.
Fig.\,\ref{fig:MeVgap} shows a composition of the CPB between $\SI{20}{keV}$ and $\SI{40}{GeV}$, which include the above sources and their combined contributions. 
Additionally, the IC component of the positron and electron energy loss spectrum is shown.
The corresponding shaded band represents the uncertainty range obtained by varying the IC contribution.
The range of the uncertainty band is derived from observational uncertainties of the COB/CUV spectrum and the CGB, which are, by far, the most dominant contributors to pair production (see Fig.\,\ref{fig:pair_yield_contribution}).
The uncertainties of the respective CPB measurements are estimated to be around 70\% for the COB/CUV and 50\% for the CGB, corresponding to relative intensity variations of $0.3 - 1.7$ and $0.5 - 1.5$, respectively, motivating an uncertainty of 0.15 and 2.55 times the fiducial IC prediction.
Additionally, a factor of 2 was included to account for uncertainties of the energy loss model, which takes into account that some positrons may annihilate earlier, such as in the interstellar medium of different galaxy types.
The cooling timescale of IC of GeV--TeV positrons inside galaxies is several orders of magnitude smaller than the annihilation time scale, so that all positrons are expected to experience the full cooling function until they annihilate at almost rest \citep[e.g.,][]{Jean2009}.
Annihilation in flight is sub-dominant in any case at a level of a few per cent \citep{Das2025_AiF,Knoedlseder2025_AiF}, so that a global factor of 2 represents the most conservative estimate for the IC component, as all electrons will participate in the IC process nevertheless.
Propagating these uncertainties yields an overall normalization range between 0.075 and 5.1 times the fiducial IC prediction.
}

It becomes evident that this process leads to a sizable contribution to the CGB by Inverse Compton scattering.
Especially in the range between 1 and 100\,MeV, that is, including the least-accurately measured part, the Inverse Compton part can make up to 10--20\% of the entire CGB.
{
When added to the presumably-known-source background, the resulting total intensity approaches the observed MeV background measurements and may substantially reduce the gap in our understanding of these measurements.
As interactions of MeV photons play a relatively minor role in contributing to the cosmological pair production processes    (Fig.\,\ref{fig:pair_yield_contribution}), the CGB spectrum in the MeV range does not get considerably reduced in intensity by electron-positron pair production.
Instead, it appears that the interactions of COB/CUV photons with those of the CGB and the eventual energy loss and annihilation leads to a sizable redistribution of those photons towards MeV energies, which would, therefore, be an additional contribution.
}

The double-peaked structure originates from both, the parent particle distributions, Fig.\,\ref{fig:pair_injection_spectrum}, as well as the changing contributions of the CMB, CIB, COB, and CUB as a function of redshift.
In fact, the spectrum is multiple-peaked but due to the redshift-integration, only the two peaks around 0.1\,MeV and tens of MeV survive.
Those are directly related to the CMB and COB/CUB, respectively.
This means while the first peak is difficult to suppress given the accuracy of CMB measurements, the second peak might be more uncertain, however in both directions.
Considering the overall shape of the spectrum, the model seems to nicely follow the measurements.

We note that the final emission spectrum strongly depends on the accuracy of the CPB measurements:
The uncertainties in the absolute intensities directly propagate into the differential emissivities and consequently into the secondary emission spectrum.
Likewise, the propagation of the individual components depends on the modelling accuracy of the redshift evolution.
However, the Inverse Compton cooling power is always dominated by the CMB, which is the best understood component in the whole model.
We estimate that the Inverse Compton spectrum may be uncertain by at most one order of magnitude, given the measurement and modelling uncertainties.

\section{Conclusion and Outlook}\label{Conclusion & Outlook}
In this paper, we calculated the electron-positron pair-production rate by photons from interactions of the CPB with itself.
The CPB was approximated by a sum of 28 diluted blackbody functions spanning the electromagnetic spectrum from 10\,MHz to 1\,TeV.
To model the CPB for different redshifts, we take into account the expansion of the Universe and the relative luminosity evolution of each contributing source type.
According to our model, the resulting pair production rate increases sharply up to a redshift of $2.7$, where it reaches a value around $10^{53}\,\mathrm{e^\pm\,s^{-1}}$.
After that, it decreases again and drops to a pair production rate in the order of $10^{52}$--$10^{51}\,\mathrm{e^\pm\,s^{-1}}$.
The local ($\lesssim 50$\,kpc) pair production rate from this mechanism is only $10^{35}\,\mathrm{e^\pm\,s^{-1}}$ and therefore eight orders of magnitude smaller than the Galactic positron annihilation rate.

The cosmologically enhanced $\gamma$-$\gamma$ pair-production rate leads to a secondary emission spectrum which is dominated by Inverse Compton scattering.
Because the CMB's energy density grows with $(1+z)^4$, and because the pair creation rate is $5$--$7$ orders of magnitude stronger than at redshift $0$, the observed Inverse Compton emission peaks around 100\,keV and may furthermore explain some part of the $1$\,MeV--{$1$\,GeV} emission of the CGB, {effectively reducing the discrepancy between accounted sources and observational measurements}.
Within uncertainties of the CPB measurements, and to a lesser extent modelling uncertainties of the redshift evolution of sources, the CGB spectrum between 1--10\,MeV may even be completely explained by this process.
The total flux of positron annihilation (Ps plus in-flight) amounts to $5 \times 10^{-6}\,\mathrm{ph\,cm^{-2}\,s^{-1}}$ in the range up to 50\,GeV, corresponding to an energy flux of $10^{-12}\,\mathrm{erg\,cm^{-2}\,s^{-1}}$.
This annihilation signal is completely dominated by the Inverse Compton emission of the non-annihilating population, leading to a total flux of $20\,\mathrm{ph\,cm^{-2}\,s^{-1}}$, or $2 \times 10^{-7}\,\mathrm{erg\,cm^{-2}\,s^{-1}}$.
However, while this process is inevitable, the measurement and modelling uncertainties may not be neglected.
Model improvements require a collaboration across the entire astronomical community as the entire electromagnetic spectrum needs to be measured and understood precisely.
The measurement and modelling uncertainties directly impact the resulting Inverse Compton spectrum, so that a systematic uncertainty in the entire band considered here of about one order of magnitude is estimated.

While this work was intended as an investigation for another possible source of positrons in the Milky Way, it provided instead a secondary source of positrons above 1\,GeV up to 1\,TeV, depending on the radiation fields considered.
In addition, it was shown that the secondary and tertiary emission of electrons and positrons in any astrophysical environment should never be ignored as they can provide strong bounds with existing data.
Finally it should be noted that positrons undergoing mainly plasma and Inverse Compton losses can indeed show in-flight annihilation peaks beyond several tens of MeV \emph{and} lead to thermalized positrons forming Ps, even in a heavily diluted medium.

\begin{acknowledgements}
      Saurabh Mittal acknowledges support by the \emph{Bundesministerium für Wirtschaft und Energie via the Deutsches Zentrum für Luft- und Raumfahrt (DLR)} under Contract No. 50\,OO\,2219. Laura Eisenberger acknowledges support by the \emph{Bundesministerium für Wirtschaft und Energie via the Deutsches Zentrum für Luft- und Raumfahrt (DLR)} under Contract No. 50\,OR\,2413 and is grateful for the support of the \emph{Stu\-di\-en\-stif\-tung des Deu\-tschen Vol\-kes}. Dimitris Tsatsis acknowledges support from the \emph{DFG/LIS} project SI\,2502/6-1, project number 551127478. Rudi Reinhardt acknowledges support by the \emph{Bun\-des\-mi\-nis\-te\-ri\-um für Forschung, Technologie und Raumfahrt} as part of the \emph{FORKA (Forschung für den Rückbau kerntechnischer Anlagen)} funding program under the reference number 15S9455\,A-E.
\end{acknowledgements}

\bibliographystyle{aa}
\bibliography{Bibliography}

%\appendix
%
%\section{Evolution Functions and Parameters for the Cosmic Photon Background}\label{app:evolution}
%

%\section{Cosmological Energy Loss Functions}\label{app:cosmo_losses}
%

\end{document}